# Towards sodium combustion modelling with liquid water

Damien Furfaro [(1)], Richard Saurel [(1,2)], Lucas David [(2,3)] and François Beauchamp [(3)]

[(1)] RS2N SAS, Saint Zacharie, France
[(2)] Aix Marseille Univ, CNRS, Centrale Marseille, LMA, Marseille, France
[(3)] CEA Cadarache, LTRS, Saint-Paul-lez-Durance, France

**Abstract**

Solid and liquid sodium combustion with liquid water occurs through a thin gas layer where exothermic reactions happen with sodium and water vapors. It thus involves multiple interfaces separating liquid and gas in the presence of surface tension, phase transition and surface reactions. The gas phase reaction involves compressible effects resulting in possible shock wave appearance in both gas and liquid phases. To understand and predict the complexity of sodium combustion with water a diffuse interface flow model is built. This formulation enables flow resolution in multidimension in the presence of complex motion, such as for example Leidenfrost-type thermo-chemical flow. More precisely sodium drop autonomous motion on the liquid surface is computed. Various modelling and numerical issues are present and addressed in the present contribution. In the author's knowledge, the first computed results of such type of combustion phenomenon in multidimensions are presented in this paper thanks to the diffuse interface approach. Explosion phenomenon is addressed as well and is reproduced at least qualitatively thanks to extra ingredients such as turbulent mixing of sodium and water vapors in the gas film and delayed ignition. Shock wave emission from the thermo-chemical Leidenfrost-type flow is observed as reported in related experiments.

**emails :**
**damien.furfaro@rs2n.eu**
**richard.saurel@univ-amu.fr**
**lucas.david@cea.fr**
**francois.beauchamp@cea.fr**

# 1. Introduction

Fast-neutron reactors (FNR) as well as other engineering systems use sodium ($N_a$) as coolant fluid. Sodium presents excellent physical properties regarding heat transfer efficiency as well as its ability to maintain kinetic energy of fast neutrons. However, it has major drawbacks regarding safety issues as it reacts exothermically with both air and water. In the limit, explosion may occur resulting in shock wave propagation in the liquid and surrounding media.

When a liquid or solid sodium drop is set on a liquid water surface surprising phenomenon occurs. A reaction appears rapidly resulting in autonomous drop motion on the liquid surface. It seems that the drop is separated from liquid water by a small gas layer where combustion occurs both in the gas phase and at sodium surface. The phenomenon is reminiscent of the Leidenfrost effect except that the heat needed to vaporize sodium and water comes from the combustion of themselves or their vapors. This combustion induces liquid water evaporation and heating of the sodium drop. After some delay, typically a few seconds, explosion occurs. These complex events are qualitatively reported on many videos available on the web, such as [www.youtube.com/watch?v=ODf_sPexS2Q] for example. They clearly illustrate complexity of the physics and chemistry in presence.

Quantitative analytic experiments have been carried out at CEA Cadarache, France, in the facilities SOCRATE, DINAMO, VIPERE and LAVINO (Carnevali 2012, Carnevali et al. 2013, Daudin 2015, Daudin et al. 2018, David et al. 2019).

These experiments confirm an important fact: liquid sodium and liquid water are always separated by a gas layer or bubbly zone, resulting in significant lowering of the energy release efficiency compared to the theoretical one. To be more precise, typical reactions of sodium with water assume (ideal) molecular mixing, resulting in energy release of the order of 100 kJ/mol, which is considerable. However, in the experiments and engineering situations of interest, sodium is never mixed with water at molecular scale. Materials are separated by interfaces and the gas layer repels the reactive material (Na) from the oxidizer (water). It results in low energy release rate, with mechanical consequences and blast effects much lower than if the reaction was occurring with molecularly mixed materials. In the limit, reactive materials never mix because of projections, resulting in incomplete reaction, with moderate energy release.

The present work attempts to model these effects to understand and predict the physics occurring in this complex two-phase combustion system. To determine the effective energy release and its kinetics in situations relevant to FNR safety, the mixing process between combustible sodium and oxidizer (water) must be modelled. Mixing of reactants controls the energy release rate. This mixing seems to occur through the gas layer separating the two liquids, resulting in temperature and pressure rise through exothermic reactions occurring in the gas phase and at the sodium surface, resulting in turn to autonomous drop motion and possibly explosion.

In the author's knowledge, the present paper is the first attempt to model sodium combustion with liquid water in multidimensional configuration. Former contributions considered multidimensional water vapor flow interacting with a liquid sodium surface at rest through a diffusion flame (Deguchi et al., 2015). Marfaing (2014) and Marfaing et al. (2014) considered both liquid and vapor water in the presence of a diffusion flame in 1D spherical configuration. The assumption of 1D flow seemed restrictive, as in reality gas escapes from the film to the atmosphere, as illustrated in the Figure 1.

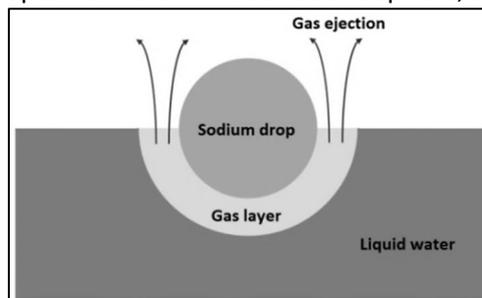

Figure 1: Schematic representation of a floating sodium drop at the surface of liquid water separated by a gas layer. The gas layer thickness is controlled by gas ejection, density difference between the liquids, diffusion fluxes in the gas layer, phase change at interfaces and exothermicity of the reactions, at sodium surface and in the gas mixture.

In this context, gas layer thickness determination is very important as it controls heat and mass diffusion fluxes between sodium and water, responsible for reactants gas mixing and reaction. As illustrated in Figure 1, self-selection of the gas layer thickness is directly linked to multidimensional effects. A 1D approach can difficulty tackle this problem and has another restriction, such as liquid caustic soda layer appearance, resulting in extra separation of reactants. It seems that the main motivation of 1D computations relied in simplicity to compute material interfaces as they can be explicitly tracked. Marfaing et al. (2014) used an Arbitrary Lagrangian Eulerian (ALE) formulation in this aim. The present approach is based on diffuse interface formulation (see Saurel and Pantano (2018) for a review). In this frame, interface deformations can be arbitrarily large as they are captured, as shocks and contact surfaces are captured in conventional gas dynamics computations. Moreover, diffuse interface approaches are able to address the complex physics and chemistry needed to model sodium-water reaction. Indeed, the relevant phenomena to address are:
- Fluids compressibility and the possible presence of shock waves;
- Presence of interfaces with complex physics, such as compressible effects, surface tension, phase transition, chemical species and thermal diffusion as well as surface reactions;
- Intense motion, the interfaces being mobile and deformable;
- Chemical reactions in the gas phase.

As thermal diffusion is present, as argued in Le Martelot et al. (2014) and Saurel et al. (2016) a single temperature diffuse interface model is relevant. Such two-phase model is reminiscent of the reactive Euler equations widely used in combustion, but with thermodynamic closure significantly different of the Dalton's law (Chiapolino et al., 2017). From this model, extra physics has to be included, such as the various chemical reactions and multiple phase transitions as both liquid water and sodium evaporate. It seems that a diffuse interface approach was considered as well by Aksenova et al. (2017) and Chudanov et al. (2019) to address the same topic of sodium-water reaction.

Diffuse interfaces methods also have limitations, the most relevant one being the excessive numerical diffusion of material interfaces. Thanks to the sharpening method of Chiapolino et al. (2017), based on a specific gradient limiter embedded in the MUSCL scheme (Van Leer, 1979), interfaces are captured with almost 2-3 mesh points at any time, showing significant improvement of solutions. Another difficulty is related to the treatment of thermal and molecular diffusion with numerically diffuse interfaces. This issue is addressed in the present paper. Another issue is related to the correct numerical treatment of capillary forces and curvature computation, that are challenging when interfaces are sharpened. This issue is addressed as well.

The present paper is organized as follows. The multi-D multiphase hydrodynamic model in velocity, pressure and temperature equilibrium, augmented with various physicochemical effects is presented in Section 2. The thermodynamic closure is specified in Section 3. Both surface and gas chemical reactions are detailed in Section 4. The interface sharpening method of Chiapolino et al. (2017) is adapted to the present framework in Section 5. In Section 6, treatment of mass diffusion with numerically diffuse interfaces is addressed. Effects of the various physical effects (phase transition, heat and mass diffusion, surface and volume chemical reactions) are illustrated in 1D spherical tests. Gravity and capillary effects are then addressed for 2D computations. Specific Riemann solvers are developed in Sections 7 and 8 respectively. Thanks to these ingredients 2D results are presented in Section 9 showing efficient computations of thermochemical Leidenfrost-type effect. To reproduce shock wave emission and explosion, extra ingredients are needed, such as turbulent mixing in the gas layer and chemical kinetics of gas phase reaction. These effects are considered in Section 10 and related computations show multidimensional explosions with shock wave emission, as reported in experiments.

**2. Multiphase model and properties**
With diffuse interface approaches, the entire domain is considered as a multiphase mixture and interfaces correspond to zones where mixture density, mass and volume fractions become discontinuous. The model considered hereafter is an extension of the one of Le Martelot et al. (2014)

and Saurel et al. (2016) developed to model phase change at interfaces. Relevant literature in this frame may be found in Kapila et al. (2001), Saurel et al. (2008), Grove (2010), Lund (2012) to cite a few. In this approach, each phase is considered compressible and governed by an appropriate equation of state (EOS), as examined in Section 3. The multiphase mixture evolves in velocity, pressure and temperature equilibrium. Temperature equilibrium is justified in the present context as heat diffusion is considered resulting in temperature continuity at interfaces.

Five chemical species are present in the sodium-water reaction context, present in two thermodynamic phases: liquid and vapor water, liquid and vapor sodium, liquid soda, hydrogen and nitrogen. In the present model, hydrogen combustion is not considered as sodium-water explosions have been observed in the absence of such reaction (Carnevali et al., 2013) where the atmosphere was made of Argon.

The corresponding flow model reads:

$$\begin{cases} \dfrac{\partial \rho Y_{H_2O}^L}{\partial t} + \text{div}\left(\rho Y_{H_2O}^L \vec{u}\right) = \rho \nu_{H_2O}\left(g_{H_2O}^g - g_{H_2O}^L\right) \\ \dfrac{\partial \rho Y_{Na}^L}{\partial t} + \text{div}\left(\rho Y_{Na}^L \vec{u}\right) = \rho \nu_{Na}\left(g_{Na}^g - g_{Na}^L\right) - \dot{\omega}_{SR} \\ \dfrac{\partial \rho Y_{NaOH}^L}{\partial t} + \text{div}\left(\rho Y_{NaOH}^L \vec{u}\right) = \varphi_{NaOH}\left(\dot{\omega}_{GR} + \dot{\omega}_{SR}\right) \end{cases} \text{(liquid phase masses)}$$

$$\begin{cases} \dfrac{\partial \rho Y_{H_2O}^g}{\partial t} + \text{div}\left(\rho Y_{H_2O}^g \vec{u} + \alpha_g \overrightarrow{F_{H_2O}^g}\right) = -\rho \nu_{H_2O}\left(g_{H_2O}^g - g_{H_2O}^L\right) - \varphi_{H_2O}\left(\dot{\omega}_{GR} + \dot{\omega}_{SR}\right) \\ \dfrac{\partial \rho Y_{Na}^g}{\partial t} + \text{div}\left(\rho Y_{Na}^g \vec{u} + \alpha_g \overrightarrow{F_{Na}^g}\right) = -\rho \nu_{Na}\left(g_{Na}^g - g_{Na}^L\right) - \dot{\omega}_{GR} \\ \dfrac{\partial \rho Y_{H_2}^g}{\partial t} + \text{div}\left(\rho Y_{H_2}^g \vec{u} + \alpha_g \overrightarrow{F_{H_2}^g}\right) = \varphi_{H_2}\left(\dot{\omega}_{GR} + \dot{\omega}_{SR}\right) \\ \dfrac{\partial \rho Y_{Air}^g}{\partial t} + \text{div}\left(\rho Y_{Air}^g \vec{u} + \alpha_g \overrightarrow{F_{Air}^g}\right) = 0 \end{cases} \text{(gas phase masses)} \quad (2.1)$$

$$\dfrac{\partial \rho \vec{u}}{\partial t} + \text{div}\left(\rho \vec{u} \otimes \vec{u} + p\mathbf{I}\right) = \rho \vec{g} + \sigma_{Na} \kappa_{Na} \overrightarrow{\nabla \alpha_{Na}^L} + \sigma_{H_2O} \kappa_{H_2O} \overrightarrow{\nabla \alpha_{H_2O}^L} \quad \text{(mixture momentum)}$$

$$\dfrac{\partial \rho E}{\partial t} + \text{div}\left[(\rho E + p)\vec{u} + \alpha_g \overrightarrow{q_M} + \overrightarrow{q_T}\right] = \rho \vec{g}.\vec{u} + \sigma_{Na} \kappa_{Na} \overrightarrow{\nabla \alpha_{Na}^L}.\vec{u} + \sigma_{H_2O} \kappa_{H_2O} \overrightarrow{\nabla \alpha_{H_2O}^L}.\vec{u} \quad \text{(mixture energy)}$$

The flow model being quite sophisticated, some details are needed. Let us consider first mass balance equations. The mass balance equation of water vapor is a relevant candidate as the various considered effects are present:

$$\dfrac{\partial \rho Y_{H_2O}^g}{\partial t} + \text{div}\left(\rho Y_{H_2O}^g \vec{u} + \alpha_g \overrightarrow{F_{H_2O}^g}\right) = -\rho \nu_{H_2O}\left(g_{H_2O}^g - g_{H_2O}^L\right) - \varphi_{H_2O}\left(\dot{\omega}_{GR} + \dot{\omega}_{SR}\right)$$

- $\rho$ ($= \sum_{k=1}^{N} \rho Y_k$ with N = 7) denotes the two-phase mixture density. $Y_{H_2O}^g$ represents the mass fraction of water vapor with respect to the two-phase mixture;
- $\vec{u}$ represents the mixture center of mass velocity;
- $\alpha_g$ represents the gas volume fraction. It is defined by $\alpha_g = \rho \sum_{k=4}^{N} \dfrac{Y_k}{\rho_k(p,T)}$, where p and T represent the mixture pressure and temperature respectively. The relation between $\rho_k$, p and T is detailed in Section 3.
- $\overrightarrow{F_{H_2O}^g}$ represents the molecular diffusion flux of water vapor. For a given chemical species k present in the gas phase it reads,

$$\vec{F_k} = C\dfrac{1}{p}\left(y_k \overrightarrow{\nabla p} - \overrightarrow{\nabla p_k}\right),$$

where C represents the diffusion coefficient, $y_k$ the mass fraction of species k in the gas mixture ($y_k = Y_k / Y_g$, with $Y_g = 1 - Y_{H_2O}^L - Y_{Na}^L - Y_{NaOH}^L$ the gas mixture mass fraction with respect to the two-phase mixture) and $p_k$ the partial pressure. Details regarding this modeling of molecular diffusion effects are given in Appendix A. As molecular diffusion is considered in the gas phase only, the diffusion flux $\overrightarrow{F_{H_2O}^g}$ is weighted by the volume fraction $\alpha_g$ in the mass balance equation;

- $\rho \nu_{H_2O}(g_{H_2O}^g - g_{H_2O}^L)$ represents water liquid-vapor mass transfer, evaporation or condensation (Saurel et al., 2008). The relaxation parameter $\nu_{H_2O}$ controls the rate at which thermodynamic equilibrium is reached, i.e., when the Gibbs free energies become equal: $g_{H_2O}^g = g_{H_2O}^L$ ($g_k = h_k - Ts_k$, where g, h and s denote respectively the specific Gibbs free energy, the enthalpy and entropy of species k). In the present computations the various relaxation parameters $\nu_k$ related to phase change are assumed to tend to infinity, meaning that local thermodynamic equilibrium is assumed. Specific thermochemical solvers based on Chiapolino et al. (2017) are used in this aim. This solver is summarized in Appendix B.
- $\dot{\omega}_{GR}$ and $\dot{\omega}_{SR}$ represent the chemical production rates respectively related to gas reaction (GR) and surface one (SR). Their modeling is detailed in Section 4. $\varphi_{H_2O}$ represents a weight factor detailed in the same section.

Let us now examine the momentum equation:

$$\frac{\partial \rho \vec{u}}{\partial t} + \text{div}(\rho \vec{u} \otimes \vec{u} + p\mathbf{I}) = \rho \vec{g} + \sigma_{Na} \kappa_{Na} \overrightarrow{\nabla \alpha_{Na}^L} + \sigma_{H_2O} \kappa_{H_2O} \overrightarrow{\nabla \alpha_{H_2O}^L}$$

In addition to the conventional Euler equation of compressible fluids, gravity effects have been added through $\rho \vec{g}$ where $\vec{g}$ denotes the gravity acceleration. Capillary forces have been added as well and are present at the liquid water-gas interface through the term $\sigma_{H_2O} \kappa_{H_2O} \overrightarrow{\nabla \alpha_{H_2O}^L}$ and at the liquid sodium-gas interface $\sigma_{Na} \kappa_{Na} \overrightarrow{\nabla \alpha_{Na}^L}$, with $\sigma_{H_2O}$ and $\sigma_{Na}$ the surface tension coefficients taken constant in the present study ($\sigma_{H_2O} = 0.07 \text{N/m}$ and $\sigma_{Na} = 0.2 \text{N/m}$). The volume fractions are defined by

$$\alpha_{H_2O}^L = \frac{\rho Y_{H_2O}^L}{\rho_{H_2O}^L(p,T)} \text{ and } \alpha_{Na}^L = \frac{\rho Y_{Na}^L}{\rho_{Na}^L(p,T)}.$$ The local interface curvatures are denoted by $\kappa_{H_2O}$ and $\kappa_{Na}$.

Surface tension effects will be examined deeper in the numerical section. This modelling corresponds to the Brackbill et al. (1992) approach, extended to compressible fluids in Perigaud and Saurel (2005). The last equation of System (2.1) corresponds to the balance energy for the two-phase mixture. The total energy is denoted by E ($E = e + \frac{1}{2}|\vec{u}|^2$, with $e = \sum_{k=1}^{N} Y_k e_k$ the mixture internal energy). In addition to conventional terms in compressible fluids, the right-hand side involves the power of the forces appearing in the right-hand side of the mixture momentum equation. The energy flux is augmented by heat diffusion effects ($\overrightarrow{q_T} = -\left(\sum_{k=1}^{3}(\alpha_k \lambda_k) + \alpha_g \overline{\lambda_g}\right)\overrightarrow{\nabla T}$), where $\lambda_k$ represents the thermal conductivity of the phase or species k, given in the Table 2.1. The gas phase thermal conductivity is defined by $\overline{\lambda_g} = \frac{1}{2}\left(\sum_{k>3} x_k \lambda_k + 1/\sum_{k>3}(x_k/\lambda_k)\right)$ (Kee et al., 1989), where $x_k$ represents the molar fraction of the species k in the gas mixture,

$$x_k = \frac{y_k/W_k}{\sum_{i>3} y_i/W_i},$$

where $W_k$ represents the molar mass of the species k.

| | Liquid water | Liquid sodium | Liquid soda | Water vapor | Sodium vapor | Hydrogen | Air |
|---|---|---|---|---|---|---|---|
| $\lambda_k \text{ (w/m/K)}$ | 0.6071 | 70 | 0.68 | $16 \times 10^{-3}$ | $45 \times 10^{-3}$ | $187 \times 10^{-3}$ | $28.2 \times 10^{-3}$ |

Table 2.1: Thermal conductivities of the seven fluids considered for the SWR modelling.

Moreover, the energy flux is augmented by heat diffusion due to mass diffusion in the gas phase ($\vec{q}_M = \sum_{k>3} h_k \vec{F}_k$ weighted by the volume fraction $\alpha_g$).

The model satisfies the fundamental principles of physics such as mixture mass, mixture momentum and mixture energy conservation. The model is also thermodynamically consistent, i.e. it satisfies the second law of thermodynamics,

$$\frac{\partial \rho s}{\partial t} + \text{div}\left(\rho s \vec{u} + \alpha_g \sum_{k>3} s_k \vec{F}_k - \frac{\lambda_c}{T}\overrightarrow{\nabla T}\right) = \begin{pmatrix} \frac{\rho \nu_{H_2O}(g^g_{H_2O} - g^L_{H_2O})^2}{T} + \frac{\rho \nu_{N_a}(g^g_{N_a} - g^L_{N_a})^2}{T} \\ + \frac{\alpha_g p}{TC} \sum_{k>3}\left(v_k |\vec{F}_k|^2\right) + \frac{\lambda_c}{T^2}|\overrightarrow{\nabla T}|^2 \\ - \frac{|\dot{\omega}_{RS}|}{T}\underset{<0}{\Delta G^0_{SR}} - \frac{|\dot{\omega}_{RG}|}{T}\underset{<0}{\Delta G^0_{GR}} \end{pmatrix} \geq 0 ,$$

where $s = \sum_{k=1}^{N} Y_k s_k$ represents the mixture entropy and $\Delta G^0$ represents the Gibbs free energy production, for a given reaction.

In the absence of diffusive effects, source and capillary terms, it can be shown easily that the system is hyperbolic with wave speeds $u$, $u-c$, $u+c$. The mixture sound speed definition is given in Le Martelot et al. (2014). Note that its precise knowledge is useless as it is always slightly lower than the Wood (1930) sound speed given by,

$$\frac{1}{\rho c_w^2} = \sum_{k=1}^{N} \frac{\alpha_k}{\rho_k c_k^2} ,$$

useful for numerical resolution of the system.

Last, the equations are Galilean invariant. The present model dealing with mixtures in mechanical and thermal equilibrium is suitable for the computation of interfaces when heat conduction is present and related boundary layers resolved, as will be done in the computations. In the absence of non-condensable gases, chemical reactions and mass diffusion, the compressible flow model of Le Martelot et al. (2014) for boiling flows is recovered.

## 3. Thermodynamic closure

The mixture equation of state results of the following algebraic system:

$T = T_k, \forall k \in \{1,...,N\}$

$v = 1/\rho = \sum_{k=1}^{N} Y_k v_k(p,T)$ (3.1)

$p = p_k, \forall k \in \{1,...,N\}$

$e = \sum_{k=1}^{N} Y_k e_k(p,T)$

In this system, fluids are in pressure and temperature equilibrium, but each one occupies its own volume or its own volume fraction. This is very different of the well-known Dalton's law, used for gas mixtures, which assumes that each gas species occupies the entire volume and that the pressure is the sum of the partial pressures:

$T_g = T_k \ ; \ V_g = V_k \ ; \ p_g = \sum_{k=1}^{N} p_k \ ; \ e_g = \sum_{k=1}^{N} Y_k e_k$ .

For ideal gases, each molecule is free to move throughout the entire volume. But for a liquid/gas mixture phases cannot occupy the whole volume.

In the context of the so-called "separate phases", the mixture EOS is a consequence of System (3.1). It therefore requires knowledge of the EOS governing each phase, to express $v_k(p,T)$ and $e_k(p,T)$.

### 3.1. Equations of state of the pure phases

Each fluid is thermodynamics is considered through the "Noble-Abel – Stiffened-Gas" (NASG) EOS (Le Metayer and Saurel, 2016). Its caloric formulation reads,

$$p_k(v_k, e_k) = \frac{(\gamma_k - 1)(e_k - q_k)}{(v_k - b_k)} - \gamma_k p_{\infty,k}.$$

The term $(\gamma_k - 1)(e_k - q_k)$ represents thermal agitation while $(v_k - b_k)$ represents short range repulsive effects. The term $\gamma_k p_{\infty,k}$, present in liquids and solids, corresponds to the attractive effects responsible of condensed matter cohesion.

The associated thermal equation of state is built from Maxwell's rules. It reads,

$$T_k(v_k, p_k) = \frac{(v_k - b_k)(p_k + p_{\infty,k})}{(\gamma_k - 1)c_{v,k}}.$$

For a given phase k, determination of the parameters $\gamma_k$, $b_k$, $p_{\infty,k}$, $c_{v,k}$ and $q_k$ is based on the experimental phase diagram. The coefficients of a liquid phase and its vapor are coupled in order to reproduce the experimental saturation curves, latent heat and saturation pressure following the method given in Le Metayer and Saurel (2016). Limitations appear in the vicinity of the critical point but this is not relevant in the SWR context.

The NASG parameters of the different fluids are listed in the Table (3.1).

|  | $\gamma_k$ | $b_k \, (m^3/kg)$ | $p_{\infty,k} \, (Pa)$ | $c_{v,k} \, (J/kg/K)$ | $q_k \, (J/kg)$ |
|---|---|---|---|---|---|
| **Liquid water** | 1.19 | $6.61 \times 10^{-4}$ | $7028 \times 10^5$ | 3610 | $-1177788$ |
| **Liquid sodium** | 1.28 | $9.168 \times 10^{-4}$ | $7452 \times 10^5$ | 995 | $-256257$ |
| **Liquid Soda** | 1.14 | $4.55 \times 10^{-4}$ | $1500 \times 10^6$ | 1830 | $-3.974 \times 10^6$ |
| **Water vapor** | 1.47 | 0 | 0 | 955 | 2077616 |
| **Sodium vapor** | 1.62 | 0 | 0 | 250 | $4.624 \times 10^6$ |
| **Hydrogen** | 1.41 | 0 | 0 | 10160 | 0 |
| **Air** | 1.4 | 0 | 0 | 920 | 0 |
| **Soda vapor** | 1.45 | 0 | 0 | 900 | $1.712 \times 10^6$ |

Table 3.1: NASG parameters of the various fluids considered for the SWR modelling.

Parameters of liquid water and its vapor are given in Le Metayer and Saurel (2016). Regarding sodium, saturation curves are given in Fink and Leibowitz (1995) and the various EOS parameters (vapor and liquid) are determined from these data. The corresponding theoretical curves are compared to the reference ones in Figure 3.1.

The soda heat capacity at constant pressure ($c^L_{p,Na_OH} = 2100 \, J/kg/K$) is taken constant and adjusted at $T \approx 1000 \, K$ from Chase (1998). The liquid soda density as a function of temperature at atmospheric pressure is given in Daubert et al. (1994) and leads to the determination of parameters $c^L_{v,Na_OH}$, $b^L_{Na_OH}$ and $p^L_{\infty,Na_OH}$. The value of $\gamma^L_{Na_OH}$ is then obtained as $c^L_{p,Na_OH} = \gamma^L_{Na_OH} c^L_{v,Na_OH}$. Finally, it should be noted that the liquid soda reference energy $q^L_{Na_OH}$ is determined to agree with the surface reaction heat release, as detailed in Section 4.

Soda vapor thermodynamic data are given in Table 3.1 but are used only in Section 9.

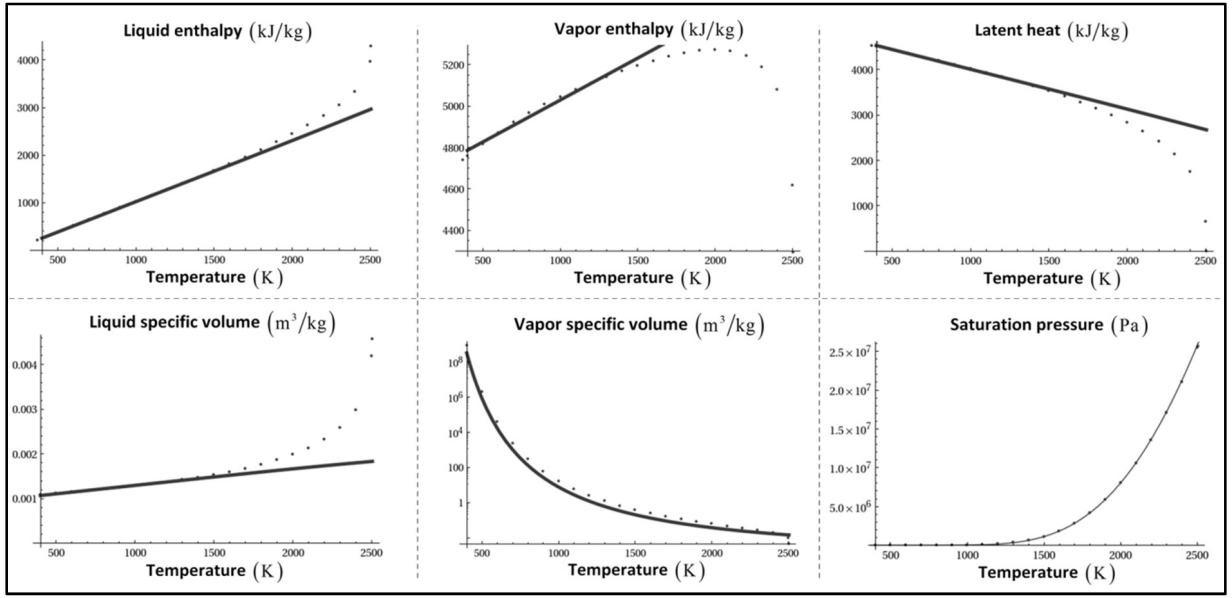
Figure 3.1: Reference (points) and theoretical (lines) sodium saturation curves. Very good agreement is observed in the temperature range $[300K - 1500K]$. This range is wide enough for SWR studies as explosions typically occur when the sodium temperature reaches about 1000K (Daudin, 2015).

### 3.2. Mixture equation of state

From System (3.1) and having in hands the NASG EOS for each phases and fluids, two relationships enable mixture temperature computation. From the mixture specific volume definition $v = \sum_{k=1}^{N} Y_k v_k(p,T)$, the mixture temperature is obtained as:

$$T(p,v) = \frac{v - \sum_{k=1}^{N} Y_k b_k}{\sum_{k=1}^{N} \frac{Y_k c_{v,k}(\gamma_k - 1)}{p + p_{\infty,k}}} \ . \tag{3.2}$$

From the mixture internal energy definition $e = \sum_{k=1}^{N} Y_k e_k(p,T)$ another relation for the mixture temperature is obtained:

$$T(p,e) = \frac{e - \sum_{k=1}^{N} Y_k q_k}{\sum_{k=1}^{N} Y_k c_{v,k} + \sum_{k=1}^{N} \frac{Y_k c_{v,k}(\gamma_k - 1)p_{\infty,k}}{p + p_{\infty,k}}} \ . \tag{3.3}$$

Combining these two relations the mixture pressure becomes solution of the equation $f(p) = 0$ with,

$$f(p) = \sum_{k=1}^{N} \frac{Y_k c_{v,k}(\gamma_k - 1)}{p + p_{\infty,k}} \left( \frac{e - \sum_{k=1}^{N} Y_k q_k}{v - \sum_{k=1}^{N} Y_k b_k} - p_{\infty,k} \right) - \sum_{k=1}^{N} Y_k c_{v,k} \ . \tag{3.4}$$

Its resolution requires an iterative method such as Newton's one.
The thermodynamic closure of System (2.1) being determined we now address modeling of the various chemical reactions.

### 4. Chemical reactions

System (2.1) involves both gas phase reactions through production terms $\dot{\omega}_{GR}$ and surface reactions occurring at the sodium surface $\dot{\omega}_{SR}$.

## 4.1 Surface reaction

Water vapor is produced at the liquid-gas interface and diffused through the gas layer to the sodium surface. As molecular diffusion is a slow process compared to the kinetics of the surface reaction, this one is considered instantaneous. A single global reaction between liquid sodium and water vapor is considered:

$$Na^{(L)} + H_2O^{(v)} \rightarrow NaOH^{(L)} + \frac{1}{2}H_2^{(g)} .\tag{4.1}$$

This reaction is exothermic, $\Delta H_{SR}^0 = -177 \text{kJ/mol}$.

Reaction (4.1) is expressed in mass terms:

$$1 kg_{Na^{(L)}} + \frac{W_{H_2O}}{W_{Na}} kg_{H_2O^{(v)}} \rightarrow \frac{W_{NaOH}}{W_{Na}} kg_{NaOH^{(L)}} + \frac{W_{H_2}}{2W_{Na}} kg_{H_2^{(g)}} ,$$

i.e.,

$$1 kg_{Na^{(L)}} + \varphi_{H_2O} kg_{H_2O^{(v)}} \rightarrow \varphi_{NaOH} kg_{NaOH^{(L)}} + \varphi_{H_2} kg_{H_2^{(g)}} ,$$

with $\varphi_{H_2O} = 0.78$, $\varphi_{NaOH} = 1.74$ and $\varphi_{H_2} = 0.04$.

As this reaction is instantaneous its numerical treatment is easy. The liquid sodium and water vapor are compared first:

- If the water vapor concentration is limiting $\left( \rho Y_{H_2O}^g < \varphi_{H_2O} \rho Y_{Na}^L \right)$, then the mass increment is computed as,

$$d\omega_{SR} = \frac{\rho Y_{H_2O}^g}{\varphi_{H_2O}} .$$

- Otherwise $d\omega_{SR} = \rho Y_{Na}^L$.

The various mass fractions are then updated as,

$$\left( \rho Y_{Na}^L \right)^* = \rho Y_{Na}^L - d\omega_{SR}$$

$$\left( \rho Y_{H_2O}^g \right)^* = \rho Y_{H_2O}^g - \varphi_{H_2O} d\omega_{SR}$$

$$\left( \rho Y_{NaOH}^L \right)^* = \rho Y_{NaOH}^L + \varphi_{NaOH} d\omega_{SR}$$

$$\left( \rho Y_{H_2}^g \right)^* = \rho Y_{H_2}^g + \varphi_{H_2} d\omega_{SR} ,$$

the state '*' being the post-reaction one.

Agreement with the surface reaction $\left( \Delta H_{SR}^0 = -177 \text{kJ/mol} = -7.702 \text{MJ/kg}_{Na} \right)$ heat release is achieved by adjusting the reference energy of liquid soda as detailed hereafter.

Each phase is governed by the NASG equation of state with the enthalpy defined as,

$$h_k(p,T) = c_{p,k}T + b_k p + q_k ,$$

where constants $b_{H_2O}^g$ and $b_{H_2}^g$ are zero.

With this definition the heat of reaction at atmospheric conditions $(p_0, T_0)$ reads,

$$\Delta H_{SR}^0 = \varphi_{NaOH}\left( c_{p,NaOH^{(L)}} T_0 + b_{NaOH}^L p_0 + q_{NaOH}^L \right) + \varphi_{H_2}\left( c_{p,H_2^{(g)}} T_0 + q_{H_2}^g \right) - \varphi_{H_2O}\left( c_{p,H_2O^{(g)}} T_0 + q_{H_2O}^g \right) - \left( c_{p,Na^{(L)}} T_0 + b_{Na}^L p_0 + q_{Na}^L \right)$$

$$\equiv -7.702 \text{MJ/kg}_{Na}$$

The reference energy of hydrogen is zero $\left( q_{H_2}^g = 0 \right)$. Those of both liquid sodium and water vapor have been determined in Section 3.1 to agree with the latent heats of vaporization.

The reference energy of liquid soda is consequently determined to respect $\Delta H_{SR}^0$:

$$q_{NaOH}^L = \frac{-7.702 \text{MJ/kg}_{Na} - \left( \varphi_{NaOH} c_{p,NaOH^{(L)}} + \varphi_{H_2} c_{p,H_2^{(g)}} - \varphi_{H_2O} c_{p,H_2O^{(g)}} - c_{p,Na^{(L)}} \right) T_0 - \varphi_{NaOH} b_{NaOH}^L p_0 + b_{Na}^L p_0 + \varphi_{H_2O} q_{H_2O}^g + q_{Na}^L}{\varphi_{NaOH}}$$

$$= -3.974 \times 10^6 \text{ J/kg}$$

This data corresponds to the one reported in Table 3.1.

## 4.2 Gas phase reaction

A single global gas reaction is considered between water vapor and sodium vapor. It produces liquid soda and gas hydrogen:

$$N_a^{(v)} + H_2O^{(v)} \rightarrow N_aOH^{(L)} + \frac{1}{2}H_2^{(g)} \tag{4.2}$$

The associated heat release is, $\Delta H_{GR}^0 = -281 kJ/mol$.

This reaction is considered instantaneous as before. Few details on the kinetics of this reaction are available (Takata and Yamaguchi, 2003). At this level it is assumed that this reaction is very fast compared to the other physical processes in presence, such as mass diffusion. This assumption will be reconsidered in Section 10.

Numerical treatment of this reaction follows the same lines as the previous one:

- If the water vapor concentration is limiting $\left(\rho Y_{H_2O}^g < \varphi_{H_2O} \rho Y_{N_a}^g\right)$, then $d\omega_{GR} = \frac{\rho Y_{H_2O}^g}{\varphi_{H_2O}}$,

- Otherwise, $d\omega_{GR} = \rho Y_{N_a}^g$.

The various mass fractions are then updated as,

$\left(\rho Y_{N_a}^g\right)^* = \rho Y_{N_a}^g - d\omega_{GR}$

$\left(\rho Y_{H_2O}^g\right)^* = \rho Y_{H_2O}^g - \varphi_{H_2O} d\omega_{GR}$

$\left(\rho Y_{N_aOH}^L\right)^* = \rho Y_{N_aOH}^L + \varphi_{N_aOH} d\omega_{GR}$

$\left(\rho Y_{H_2}^g\right)^* = \rho Y_{H_2}^g + \varphi_{H_2} d\omega_{GR}$,

the state '*' being the post-reaction one.

The flow model (2.1) with thermodynamic closure (3.4), complemented by mass transfer treatment given in Appendix B and chemical production rates given by equations (4.1) and (4.2) form a closed system of equations. Its numerical resolution poses many challenges that are addressed gradually, the first one being related to the capture of interfaces and associated numerical diffusion.

## 5. Interface sharpening

In this section, System (2.1) is considered in the absence of heat and mass diffusion, capillary effects and source terms. It thus reduces to the hyperbolic part of the model with thermodynamic closure (3.4). Its numerical resolution is achieved in the DALPHADT© code based on both structured and unstructured meshes [www.rs2n.eu]. The Riemann solver used to compute the intercell flux is the HLLC one (Toro et al., 1994), see Saurel et al. (2016) for adaptation to the present flow model. MUSCL type reconstruction is used, as detailed in Chiapolino et al. (2017) for unstructured grids. As detailed in this reference a well-known issue appears during the simple transport of an interface. Numerical diffusion smears interfaces over several points even when compressive limiters, such as Superbee (Roe, 1985) are used. To overcome this difficulty the Overbee limiter of Chiapolino et al. (2017) is used. It consists in a compressive limiter valid for linearly degenerate fields only and transport of Heaviside function, such as volume fraction separating two fluids.

The method is adapted to the flow model (2.1) by adding extra equations with respect to the volume fractions,

$$\frac{\partial \alpha_k}{\partial t} + \vec{u}.\overrightarrow{\nabla \alpha_k} = 0, \tag{5.1}$$

where k represents : liquid water $(H_2O)_L$, liquid sodium $(N_a)_L$, liquid soda $(N_aOH)_L$, multi-component gas $(g)$.

The Overbee limiter is used for these equations only. Denoting by $\phi_f$ the ratio of two gradients adjacent to a cell face ($\phi_f = (\alpha_{i+1} - \alpha_i)/(\alpha_i - \alpha_{i-1})$ in the simplified case of structured mesh), extrapolation from the cell center to the face is done through limited gradients through the formula:
$\theta(\phi_f) = \max[0, \min(2\phi_f, 2)]$.

This limiter is used at interfaces only, that are detected as,
$\alpha_k^n \alpha_j^n > \varepsilon$, with $j \neq k$.

According to numerical experiments, $\varepsilon = 10^{-2}$ seems a fair choice.

The other flow variables present in System (2.1) are computed with the same method, except that the Minmod limiter is used.

System (5.1) is solved during each time step of the hyperbolic solver. Sharpening volume fractions with the help of Overbee limiter results in enhanced accuracy in the various fluxes computation and consequently in sharpened mixture density and mass fractions.

But adding System (5.1) to System (2.1) corresponds to an overdetermined system. At the end of the time step all variables must be compatible and this is done through the reset of the various volume fractions. With the help of the computed equilibrium pressure $p_{eq}$ and temperature $T_{eq}$ given by (3.4) and (3.3), based on mass fractions (and not volume fractions) the volume fractions are reset as:
-For liquids,
$$\alpha_k = \frac{\rho Y_k}{\rho_k(p_{eq}, T_{eq})}, \forall k \leq 3.$$

-For the multicomponent gas mixture (made of 4 species):
$$\alpha_g = \rho \sum_{k=4}^{7} \frac{Y_k}{\rho_k(p_{eq}, T_{eq})}.$$

The method is illustrated on the advection test of Figure 5.1 with data relevant to SWR situations. Let us consider a 1D 32cm long tube with an interface separating two fluids (nearly pure liquid water on the left, nearly pure air on the right). The flow variables are uniform in the whole domain,
$$\begin{cases} u^0 = 0.1 m/s \\ p^0 = 0.1 MPa \\ T^0 = 300 K \end{cases},$$

except the mass and volume fractions, as well as mixture density that are discontinuous at x=0.05 m at initial time. Three computations are compared at the same time in Figure 5.1. The accuracy gain observed on the interface capture is noticeable with Overbee.

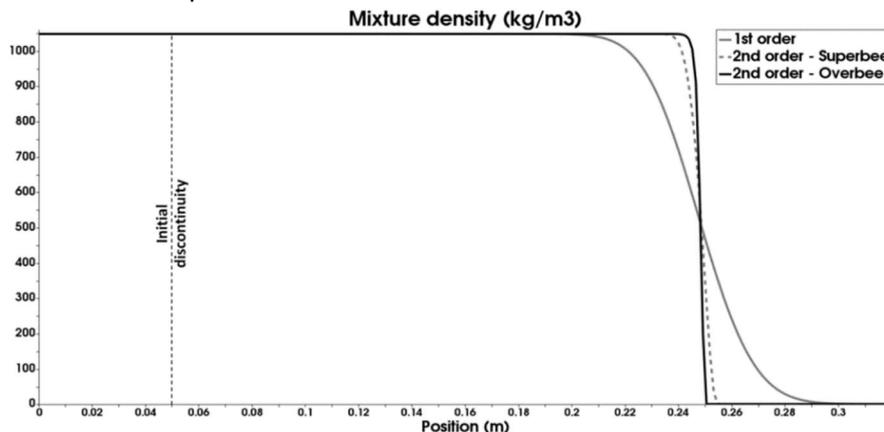

Figure 5.1: Comparison between Superbee and Overbee limiters for the advection of a liquid-gas interface at low speed – Mesh: 240 cells – CFL=0.8 – Final time: 1.98s. As the CFL is based on the sound speed more than 3 600 000-time steps are required to reach final time.

Another difficulty is now addressed and is related to the numerical treatment of mass diffusion with numerically diffuse interfaces. Indeed, even if the Overbee limiter reduces numerical diffusion, interfaces are still diffused and need special care when physical mass diffusion is considered.

# 6. Mass diffusion with diffuse interfaces and effects of chemical reactions

To illustrate the difficulty let us consider a 1D configuration relevant to SWR. A spherical liquid drop of 1mm radius is set in liquid water. The two liquids are separated by a 2 mm gas layer at elevated initial temperature. A schematic representation of the configuration under interest is shown in Figure 6.1.

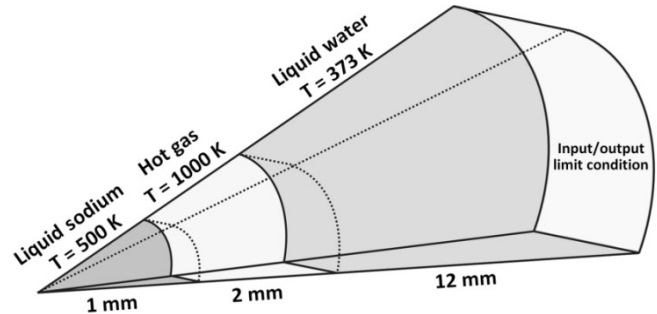

Figure 6.1: 1D configuration relevant to typical SWR situations with a 1 mm radius sodium droplet separated by a gas layer from liquid water. The liquid water domain boundary is treated as an inflow/outflow boundary condition. Precisely liquid water tank at atmospheric pressure and temperature 373 K is imposed at outflow. This outflow may become an inflow if the flow becomes inverted. Details are given in Appendix C.

For a first run, mass diffusion is removed as well as chemical reactions, both in the gas phase and at sodium surface. Surface tension and gravity effects are obviously absent. Therefore, only fluid motion is considered in the presence of heat conduction and phase transition of both water and sodium. Corresponding results are shown in Figure 6.2.

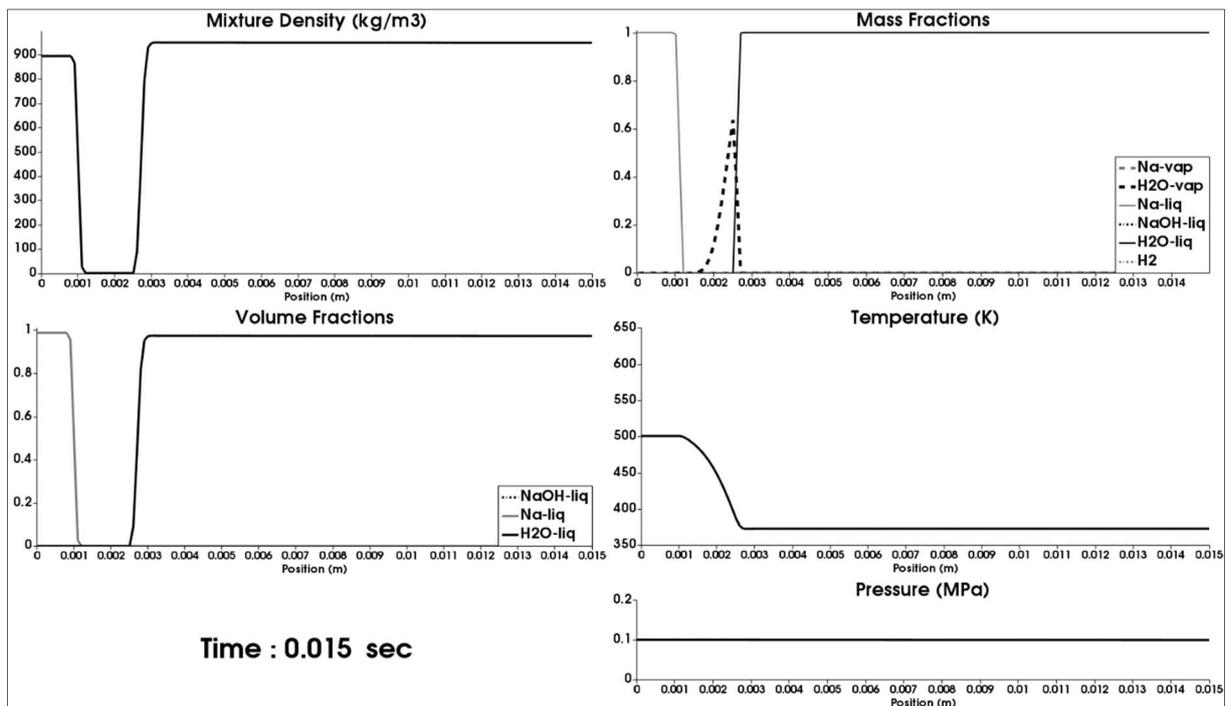

Figure 6.2: 1D reference results related to the 1D SWR test problem of Figure 6.1. Computed results are shown at time 15 ms on a mesh involving 150 cells. Thermal diffusion and phase transition only are present in the flow model (2.1). From the initial situation the gas layer has been cooled and water vapor appears at the interface at right. Both interface motions have been considered but their velocity is not significant in the present example.

In the second run, mass diffusion within the gas phase is considered with constant mass diffusion coefficient: $C = 10^{-4} \, \text{kg/m/s}$. Corresponding results are shown in Figure 6.3 at time 15ms (same as before). The water vapor created at the gas-liquid water interface is now diffused within the multi-component gas until it reaches the liquid sodium-gas interface.

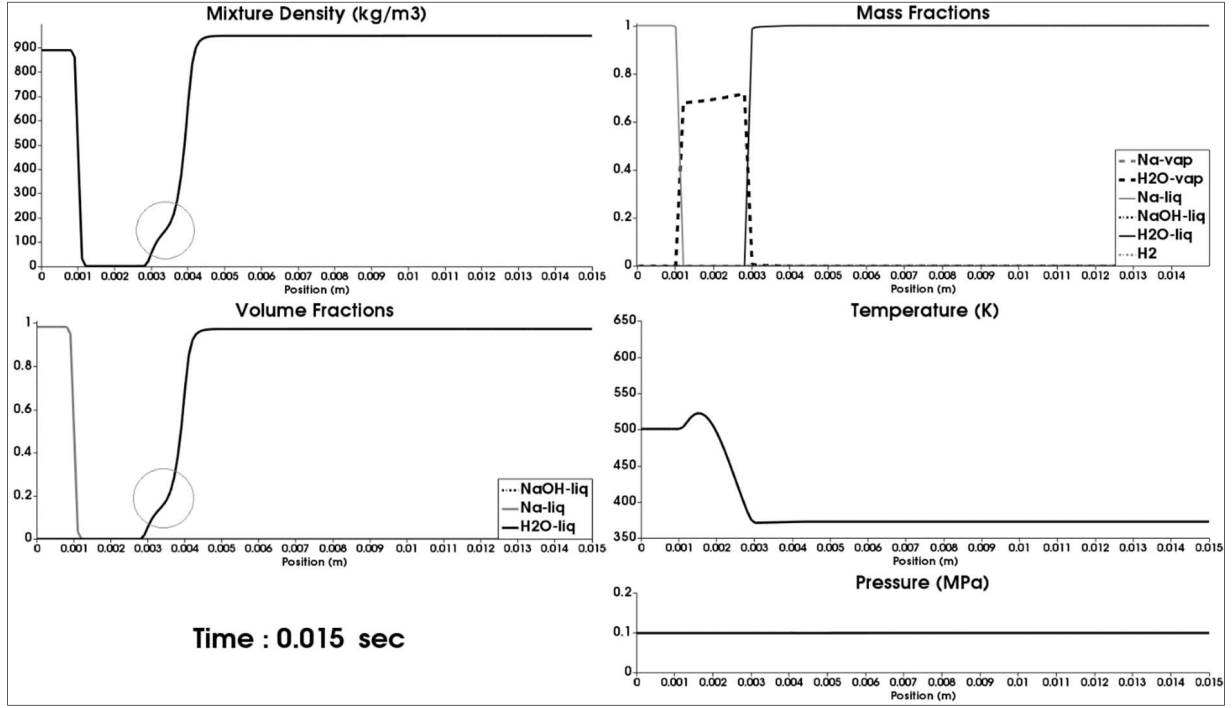

Figure 6.3: 1D computed results related to the 1D SWR test problem of Figure 6.1 in the presence of mass diffusion in addition to fluid motion, heat diffusion and phase transition already considered in the results of Figure 6.2. Same mesh is considered, and the results are shown at the same time. The gas-liquid water interface becomes corrupted as a consequence of mass diffusion in the numerically diffuse interface.

The interface separating the gas mixture and liquid water becomes unphysical, as clearly visible on the mixture density and volume fraction graphs. This issue results of bad interaction between the mass diffusion flux and the diffuse interface representation.

In the frame of diffuse interface methods, a 'pure' liquid is numerically treated as a mixture with a liquid volume fraction equal to $1-\varepsilon$. The other fluids in minor proportion then share the residue $\varepsilon$. This remark also holds for the related mass and molar fractions. In liquid water, the multi-component gas mass fraction, and in particular the water vapor one, are therefore non-zero. This implies a non-zero water vapor molar fraction (although physically inconsistent) in the liquid water domain. Computation of the related gradient $\overrightarrow{\nabla x_{H_2O}}$ present in the mass diffusion flux is consequently wrong.

Weighting the related mass diffusion flux by the gas volume fraction, as done in System (2.1) through the term $\alpha_g \overrightarrow{F^g_{H_2O}}$ is not enough to cure this problem.

To circumvent this difficulty the mass diffusion coefficient C is rendered dependent to the liquid water mass fraction as follows:

$$C\left(Y^L_{H_2O}\right) = C_0 H_\chi\left(Y^L_{H_2O}\right),\tag{6.1}$$

with $C_0 = 10^{-4}$ kg/m/s, $H_\chi\left(Y^L_{H_2O}\right) = \begin{cases} 1, & \text{if } Y^L_{H_2O} \leq 0.5+\chi \\ 0, & \text{if } Y^L_{H_2O} > 0.5+\chi \end{cases}$ and $\chi \in \,]-0.5;0.5[$ a parameter to define.

Function $H_\chi\left(Y^L_{H_2O}\right)$ is shown in Figure 6.4.

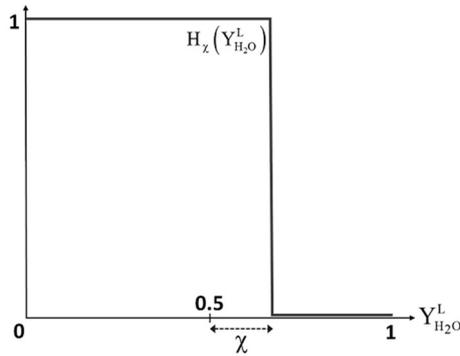

Figure 6.4: Function $H_\chi\left(Y_{H_2O}^L\right)$ used in the mass diffusion coefficient correction.

Function $H_\chi\left(Y_{H_2O}^L\right)$ prevents mass diffusion in quasi-pure liquid zones and prevents interface corruption. $H_\chi\left(Y_{H_2O}^L\right)$ is used only at the gas-liquid water interface. Indeed, at the sodium-gas interface the presence of surface reaction prevents such defect.

With the help of this non-linear diffusion coefficient, the 1D spherical test of Figure 6.1 is rerun. Parameter $\chi$ is set to $\chi = 0.2$. Results are shown at the same time as before (15ms) in Figure 6.5, showing correct interface behavior.

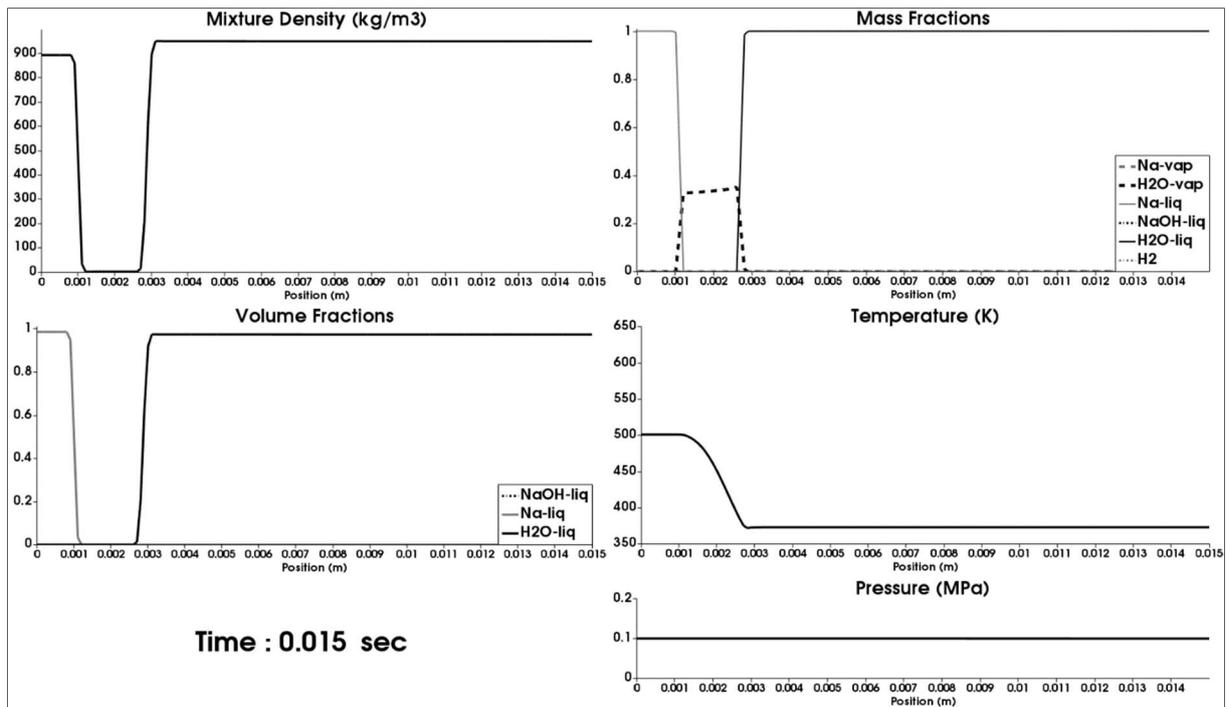

Figure 6.5: 1D computed results related to the 1D SWR test problem of Figure 6.1 in the presence of mass diffusion with non-linear mass diffusion coefficient (6.1). Fluid motion, heat diffusion, phase transition and mass diffusion are considered on the same mesh as in the previous computations, with 150 cells. Sharp liquid water - gas mixture interface is now recovered and mass diffusion in the gas phase is present. Mass diffusion effects appear clearly by comparing the mass fraction graphs of the present figure and the one of Figure 6.2.

**Influence of the parameter $\chi$**

The 1D spherical test of Figure 6.1 is rerun for different values of parameter $\chi$ in order to assess its influence on the numerical results. For each test the water vapor mass fraction at the gas-liquid water interface is recorded at different times and shown in Figure 6.6. Influence of parameter $\chi$ in the range $[-0.2; 0.2]$ appears not noticeable.

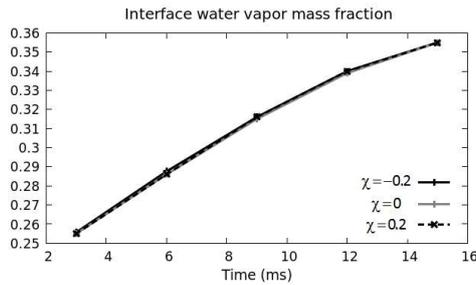

Figure 6.6: The 1D SWR test problem of Figure 6.1 is considered in the presence of fluid motion, heat diffusion, phase transition and mass diffusion. Results obtained with 3 different values of parameter $\chi$ are compared. The quantity of water vapor produced at the interface does not depend on the parameter $\chi$ in the range $[-0.2;0.2]$. Outside this range, differences appear.

The interfacial water vapor mass fraction does not depend on the parameter $\chi$ in the range $[-0.2;0.2]$ at least in the present configuration. This parameter is set to $\chi = 0.2$ in the following.

As water vapor now diffuses from the liquid water side to the sodium one it becomes possible to address both surface and volume reactions. To this end the various production terms addressed in Section 4 are used. Computed results are shown at the same time as before, on the same mesh and are shown in Figure 6.7.

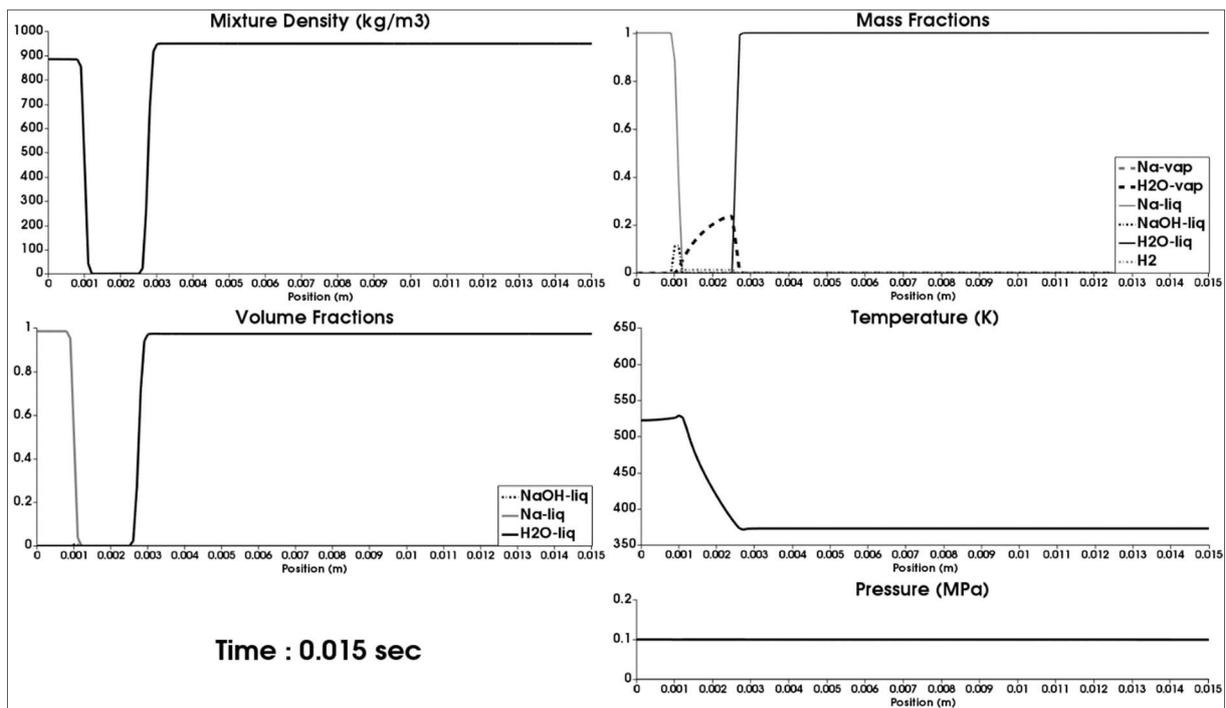

Figure 6.7: 1D computed results related to the 1D SWR test problem of Figure 6.1 in the presence of both surface and gas reactions. Fluid motion, heat diffusion, phase transition, mass diffusion $(\text{with } \chi = 0.2)$ and chemical reactions are considered on the same mesh as in the previous computations, with 150 cells. Results are shown at the same time as before (15ms). The liquid sodium temperature increases, compared to the results of Figure 6.5, as a consequence of surface reaction combined to heat diffusion.

When the water vapor reaches the liquid sodium-gas interface, surface reaction occurs, generating large soda mass production and local heating. It contributes to both liquid sodium and gas film heating through thermal diffusion. At this level, the interface motion is still imperceptible.

Results at longer times are shown in Figures 6.8 (t=1.503s) and 6.9 (t=2.334s). At time 1.503s, the liquid sodium temperature exceeds 1000K. At this temperature, a larger amount of sodium vapor is produced. This vapor then reacts with the water one through the gas reaction. The gas reaction being more exothermic than the surface one, the temperature locally exceeds 1800K. The mass fraction of the produced liquid soda is locally close to 1 and acts as a barrier between sodium and water vapors present in large quantities. The SWR gas product (hydrogen $H_2$) is diffused within the gas layer.

In addition, the gas film thickness has significantly increased, compared to the one observed in Figure 6.7, at time t=15ms.

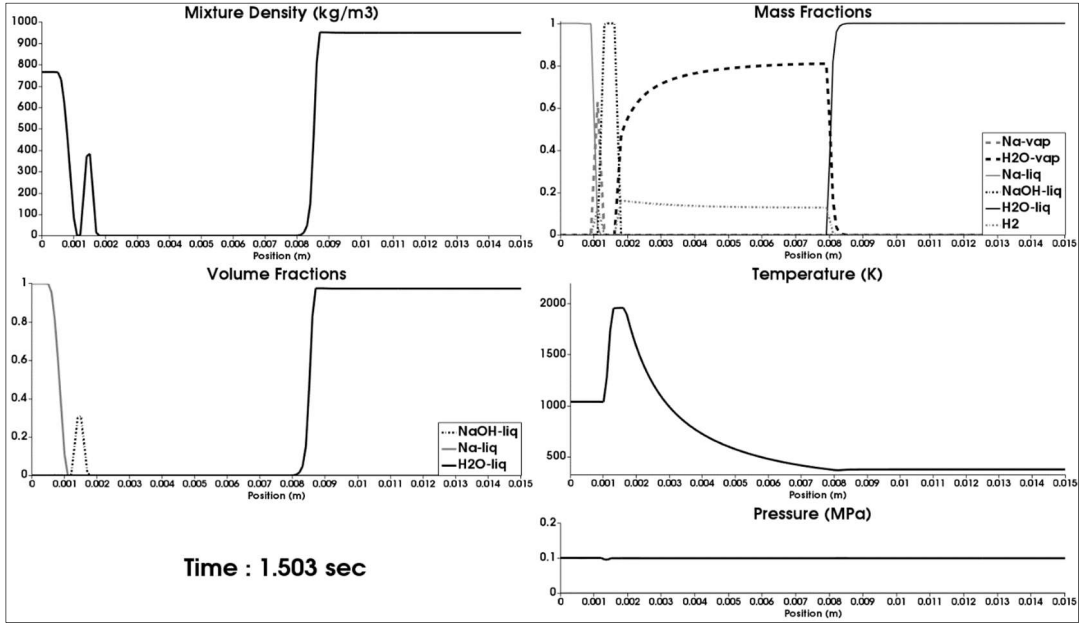

Figure 6.8: 1D computed results related to the 1D SWR test problem of Figure 6.1 in the presence of fluid motion, heat diffusion, phase transition, mass diffusion $(\chi=0.2)$ and both surface and gas reactions. The mesh made of 150 cells is still considered. Results are shown at time t=1.503s. At this stage, the gas film thickness has considerably increased. The liquid sodium temperature exceeds 1000K. The liquid soda film separates both sodium and water vapors present in large proportions.

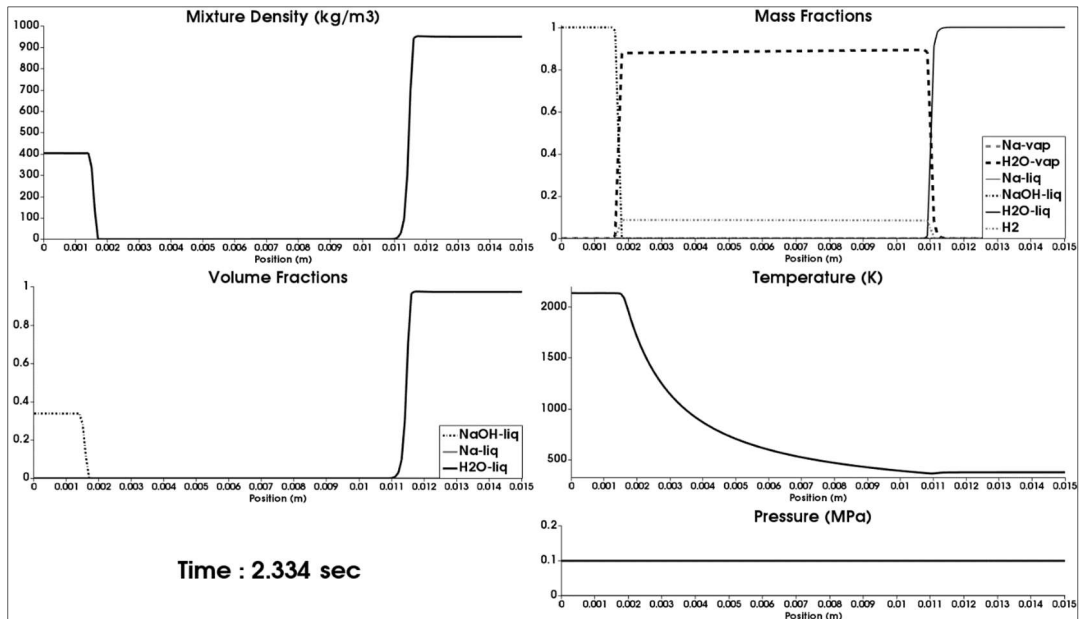

Figure 6.9: 1D computed results related to the 1D SWR test problem of Figure 6.1 in the presence of fluid motion, heat diffusion, phase transition, mass diffusion $(\chi=0.2)$ and both surface and gas reactions. The mesh made of 150 cells is considered again. Results are shown at time t=2.334s. At this stage, the sodium has been fully consumed. A hot (about 2000K) soda drop is formed at the place occupied initially by sodium.

At time 2.334s (Figure 6.9), liquid sodium has exceeded boiling point ($\approx 1156K$ at atmospheric pressure) and sodium vapor has been fully consumed. It results in hot $(\approx 2000K)$ soda drop formation. The gas film continues to grow, while the soda drop gradually cools. No explosion (with shock wave emission) is observed. The pressure is nearly uniform in the Figures 6.2 to 6.9.

Three main points emerge of the former 1D numerical experiments:
- The diffuse interface model is able to model at least qualitatively the complex physics present in SWR process;
- The combustion regime observed in these tests is governed by thermal and mass diffusion. As the distance increases monotonically versus time the various gradients and associated fluxes decrease forbidding explosion occurrence;
- A liquid soda layer appears close to the sodium surface, lowering gas mass diffusion from the liquid water surface.

These observations motivate multi-D modelling, as illustrated in Figure 6.10. Indeed, the gas layer width is selected as a consequence of the various diffusive and reactive effects, in competition with gas ejection at the free surface and weight of the sodium drop. Also, gas ejection has potential to remove liquid soda layer. Multi-D effects are expected to maintain intense gradients oppositely to 1D computations. However, two main numerical issues appear in addition to those already addressed in Sections 5 and 6. These issues are related to body and surface forces and are addressed in the forthcoming sections.

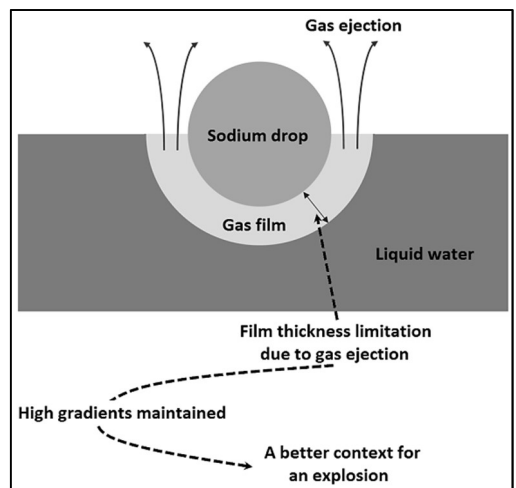

Figure 6.10: Schematic representation of the motivations for 2D computations. The gas layer width is selected by the competition of heat and mass diffusion, phase transition, exothermicity of the various reactions against gas ejection and weight of the sodium drop.

## 7. Numerical approximation of gravity effects

Accurate computation of gravity effects is of primary importance in the present context as gas layer width selection, directly linked to the various diffusive effects present in the flame, is closely driven by buoyancy.

System (2.1) involves gravity through source terms present in both mixture momentum and mixture energy balance equations. Source term splitting methods are well known to produce instabilities, particularly when dense fluids are considered. Insertion of gravity effects in the flux computation, through 'well balanced' Riemann solvers has been the subject of many efforts, such as for example LeVeque (1998), Gosse (2000) and Gallice (2002).

In the present section the HLLC solver (Toro et al., 1994) is considered and gravity effects are embedded in the formulation. This solver is genuinely positive, an important property when dealing with material interfaces and large density ratios, as well as sophisticated EOS, as the one given in Section 3.

Let us consider the hyperbolic part of System (2.1):

$$\frac{\partial U}{\partial t} + \text{div}[F] = 0, \tag{7.2}$$

with $U = \begin{pmatrix} \rho Y_k \\ \rho \vec{u} \\ \rho E \end{pmatrix}$ and $F = \begin{pmatrix} \rho Y_k \vec{u} \\ \rho \vec{u} \otimes \vec{u} + p\mathbf{I} \\ (\rho E + p)\vec{u} \end{pmatrix}$.

In the HLLC solver framework, three waves are considered:
- The extreme waves speeds $S_L$ and $S_R$, estimated as (Davis, 1988),

$$S_L = \min\left((\vec{u}.\vec{\eta_f})_L - c_L, (\vec{u}.\vec{\eta_f})_R - c_R\right)$$

$$S_R = \max\left((\vec{u}.\vec{\eta_f})_L + c_L, (\vec{u}.\vec{\eta_f})_R + c_R\right),$$

with $\vec{\eta_f}$ the unit normal vector of the face f oriented towards the cell R. For the sake of simplicity, normal velocities are denoted by $(\vec{u}.\vec{\eta_f})_L \equiv u_L$ and $(\vec{u}.\vec{\eta_f})_R \equiv u_R$ in the following.

- The contact discontinuity speed $S_M$, to determine in the presence of gravity effects.

The Rankine-Hugoniot relations through the extreme waves read,

$$(F.\vec{\eta_f})_L^* = (F.\vec{\eta_f})_L + S_L(U_L^* - U_L), \qquad (F.\vec{\eta_f})_R^* = (F.\vec{\eta_f})_R + S_R(U_R^* - U_R) \tag{7.3}$$

and lead, in particular, to the usual expressions of pressure in the star regions:

$$p_L^* = p_L + \rho_L(u_L - S_L)(u_L - S_M), \qquad p_R^* = p_R + \rho_R(u_R - S_R)(u_R - S_M) \tag{7.4}$$

In the framework of the standard HLLC solver without gravity effects, the pressure equality is ensured through the contact wave ($p_L^* = p_R^*$). But in the present context, the determination of the contact condition requires the integration of the equilibrium condition:

$$\vec{\nabla} p = \rho \vec{g},$$

with $\vec{g} = (0 \quad -g)^T$ the gravity vector.

The equilibrium condition can be rewritten as follows:

$$\begin{cases} \frac{\partial p}{\partial x} = 0 \\ \frac{\partial p}{\partial y} = -\rho g \end{cases} \tag{7.5}$$

Integration of the second relation is achieved on both sides of face f separating the two cells L and R:

$$\int_{p^*}^{p_R} dp = -\int_{\bar{y}}^{y_R} \rho g dy \Leftrightarrow p_R + \rho_R g y_R = p^* + \rho_R g \bar{y}, \qquad \int_{p_L}^{p^*} dp = -\int_{y_L}^{\bar{y}} \rho g dy \Leftrightarrow p_L + \rho_L g y_L = p^* + \rho_L g \bar{y} \tag{7.6}$$

with $p^*$ the pressure solution of the Riemann problem and $\bar{y}$ the vertical coordinate of the center of face f (Figure 7.1).

Relations (7.6) can be rewritten in the star states as follows:

$$\begin{cases} p_R^* + \rho_R g y_R = p^* + \rho_R g \bar{y} \\ p_L^* + \rho_L g y_L = p^* + \rho_L g \bar{y} \end{cases}, \tag{7.7}$$

where the density variations have been assumed to be negligible through the acoustic waves.

Combining (7.4) and (7.7) leads to the determination of both contact wave speed $S_M$ and star pressure $p^*$:

$$\begin{cases} S_M = \dfrac{p_R - p_L + \rho_L u_L (S_L - u_L) - \rho_R u_R (S_R - u_R) + \rho_R g (y_R - \bar{y}) - \rho_L g (y_L - \bar{y})}{\rho_L (S_L - u_L) - \rho_R (S_R - u_R)} \\ p^* = \dfrac{(p_R + \rho_R (u_R - S_R) u_R)\rho_L (u_L - S_L) - (p_L + \rho_L (u_L - S_L) u_L)\rho_R (u_R - S_R)}{\rho_L (u_L - S_L) - \rho_R (u_R - S_R)} + \dfrac{\rho_L (u_L - S_L)\rho_R g (y_R - \bar{y}) - \rho_R (u_R - S_R)\rho_L g (y_L - \bar{y})}{\rho_L (u_L - S_L) - \rho_R (u_R - S_R)} \end{cases} \tag{7.8}$$

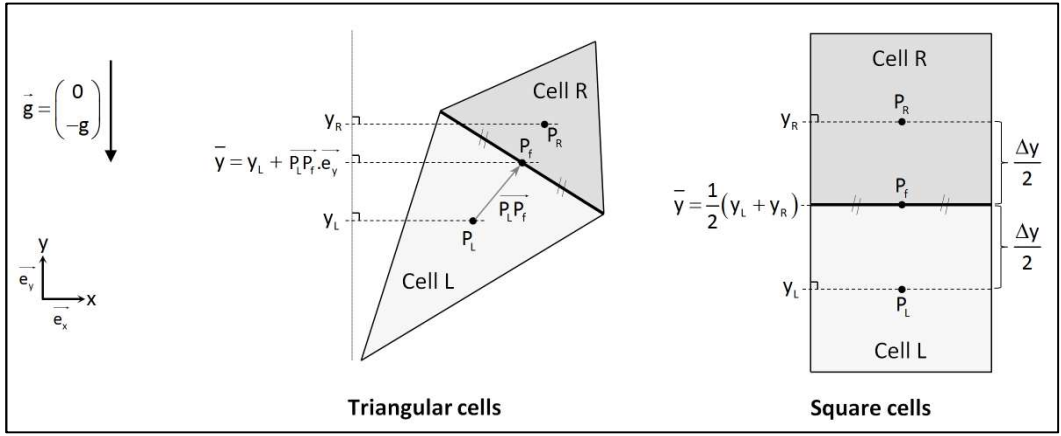

Figure 7.1: Determination of $\bar{y}$ for two mesh types (triangular and cartesian grids). $P_L$ and $P_R$ represent respectively the centers of cells L and R. $P_f$ is the center of the face f.

The primitive variable vector in the star states is thus fully determined and the solution sampling is done to compute the flux F of System (7.2).

For example, in the subsonic case such as $S_M > 0$, the flux solution of the Riemann problem on a given face f along the face normal vector $\vec{\eta}_f$ reads,

$$\left(F.\vec{\eta}_f\right)^* = \begin{pmatrix} \rho_L^* Y_{k,L}^* S_M \\ \rho_L^* S_M \vec{u}_L^* + p^* \vec{\eta}_f \\ \left(\rho_L^* E_L^* + p^*\right) S_M \end{pmatrix}, \text{ with } \begin{cases} Y_{k,L}^* = Y_{k,L} \\ \rho_L^* = \rho_L \dfrac{u_L - S_L}{S_M - S_L} \\ E_L^* = E_L + \dfrac{p_L u_L - p^* S_M}{\rho_L (u_L - S_L)} \\ \vec{u}_L^* = S_M \vec{\eta}_f + \vec{u}_L^{\perp} \end{cases}.$$

In these formulas, $\vec{u}_L^{\perp} = \vec{u}_L - u_L \vec{\eta}_f$ denotes the velocity vector tangential to the face f. Moreover, $S_M$ and $p^*$ are defined by (7.8). The use of $p^*$ instead of $p_L^*$ and $p_R^*$ given by (7.4) is a consequence of the pressure profile linearity, characteristic of gravity effects. Let us consider the specific case of a state close to the equilibrium one. Any small velocity fluctuation modifying the sign of $S_M$ would lead to an unacceptable variations between $p_L^*$ and $p_R^*$ if the latter was chosen to compute the flux $F^*$. The choice of $p^*$ therefore allows to preserve the mechanical equilibrium condition.

In the supersonic case such as $S_L > 0$, the solution reads:

$$\left(F.\vec{\eta}_f\right)^* = \begin{pmatrix} \rho_L Y_{k,L} u_L \\ \rho_L u_L \vec{u}_L + p^* \vec{\eta}_f \\ \left(\rho_L E_L + p^*\right) u_L \end{pmatrix}, \text{ with } p^* = p_L + \rho_L g \left(y_L - \bar{y}\right).$$

The following Godunov type scheme is then used to update the solution. For the sake of simplicity its formulation is given hereafter at first order,

$$U_i^{n+1} = U_i^n - \frac{\Delta t}{S_i} \sum_{f=1}^{N_{Faces}} \left(F.\vec{\eta}_f\right)^* L_f + \Delta t H_i^n,$$

$$\text{with } U = \begin{pmatrix} \rho Y_k \\ \rho \vec{u} \\ \rho E \end{pmatrix}, \ F = \begin{pmatrix} \rho Y_k \vec{u} \\ \rho \vec{u} \otimes \vec{u} + p\bar{\bar{I}} \\ (\rho E + p) \vec{u} \end{pmatrix} \text{ and } H = \begin{pmatrix} 0 \\ \rho \vec{g} \\ \rho \vec{g}.\vec{u} \end{pmatrix}.$$

$L_f$ denotes the length of the face f, $S_i$ is the surface of the cell i and $N_{Faces}$ represents the number of faces of the considered cell. The superscripts n and n+1 denote two successive time steps $t^n$ and $t^{n+1}$ such as $\Delta t = t^{n+1} - t^n$.

Efficiency of this method is illustrated on the following test case. Let's consider a 10m height tank, the lower half tank being filled with water and the upper one with air. The initial pressure in the entire vessel is the atmospheric one. The considered mesh is coarse (100 cells) to highlight differences between the conventional splitting method and the present one, where gravity effects are embedded in the Riemann solver. Corresponding results are shown in the Figure 7.2.

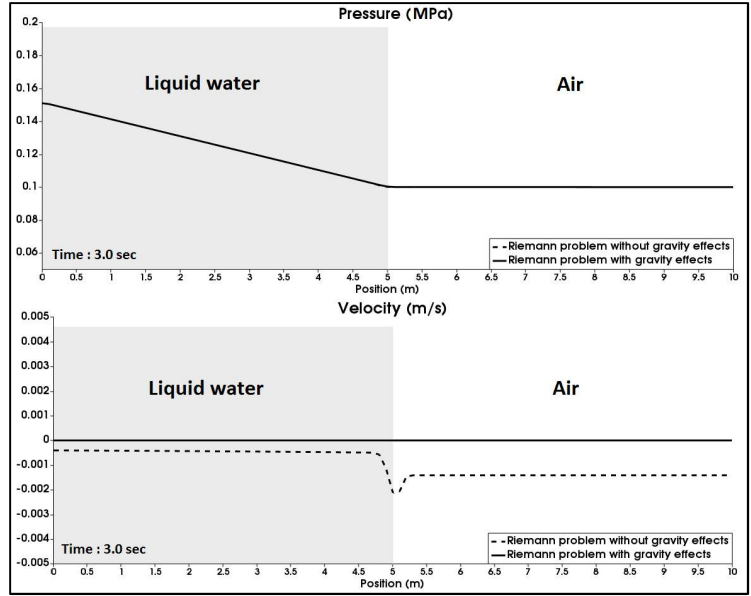

Figure 7.2: Mechanical equilibrium of a water column in air. The conventional Godunov method with source terms splitting is compared to the present one, where gravity terms are embedded in the HLLC solver. Both methods compute correct pressure field, but the new method only is free of parasitic velocity when equilibrium is reached. Computed results are shown at time 3 s.

The equilibrium state is perfectly matched with the new method. We now address surface tension effects approximation with similar approach.

## 8. Numerical approximation of surface tension

Surface tension effects are considered through the Continuum Surface Force (CSF) method of Brackbill et al. (1992). The capillary force is modelled as:

$$\vec{F_\sigma} = \sigma \kappa \overrightarrow{\nabla \alpha_{liq}},$$

where $\sigma$ represents the surface tension coefficient, $\alpha_{liq}$ is the liquid volume fraction and $\kappa$ represents the local curvature $(m^{-1})$:

$$\kappa = -\text{div}\left(\frac{\overrightarrow{\nabla \alpha_{liq}}}{|\overrightarrow{\nabla \alpha_{liq}}|}\right) \qquad (8.1)$$

In system (2.1) these effects are present through the term ($\sigma_{N_a} \kappa_{N_a} \overrightarrow{\nabla \alpha_{N_a}^L} + \sigma_{H_2O} \kappa_{H_2O} \overrightarrow{\nabla \alpha_{H_2O}^L}$) in the momentum and energy equations. Two contributions are present as two interfaces are considered.

The CSF method has been already considered with compressible fluids and diffuse interfaces (Perigaud and Saurel, 2005, Le Martelot et al., 2014, Garrick et al. 2017, Schmidmayer et al., 2017) but extra difficulties appear in the present application as a result of interface sharpening through the method presented in Section 5. Computation of the local curvature becomes problematic. The approach used in the present work is described gradually in the following.

### 8.1 Volume fraction gradient determination at cell centers

A robust and accurate method for the computation of the volume fraction gradient is based on least squares approximation. It is based on multiple Taylor expansions around cell center $P_i$ and a cloud of neighboring cells indexed by j:

$$\alpha_j = \alpha_i + \overrightarrow{P_iP_j}.\overrightarrow{e_x}\frac{\partial \alpha_i}{\partial x} + \overrightarrow{P_iP_j}.\overrightarrow{e_y}\frac{\partial \alpha_i}{\partial y} + O\left(\left\|\overrightarrow{P_iP_j}\right\|^2\right)$$
$$= \alpha_i + \Delta x_{ij}\frac{\partial \alpha_i}{\partial x} + \Delta y_{ij}\frac{\partial \alpha_i}{\partial y} + O\left(\left\|\overrightarrow{P_iP_j}\right\|^2\right) \quad , \tag{8.2}$$

where $\overrightarrow{e_x}$ and $\overrightarrow{e_y}$ denote the unit vectors of the Cartesian basis.

Using (8.2) with a set of N neighbors ($j \in \{1,...,N\}$) results in the following system:

$$\begin{pmatrix} \omega_1 \Delta x_{i1} & \omega_1 \Delta y_{i1} \\ . & . \\ . & . \\ . & . \\ \omega_N \Delta x_{iN} & \omega_N \Delta y_{iN} \end{pmatrix} \begin{pmatrix} \dfrac{\partial \alpha_i}{\partial x} \\ \dfrac{\partial \alpha_i}{\partial y} \end{pmatrix} = \begin{pmatrix} \omega_1(\alpha_1 - \alpha_i) \\ . \\ . \\ . \\ \omega_N(\alpha_N - \alpha_i) \end{pmatrix} \Leftrightarrow AX = B,$$

with $\omega_j = \dfrac{1}{\Delta x_{ij}^2 + \Delta y_{ij}^2}$.

Weights $\omega_j$ allows to control numerical instabilities (division by small numbers) when the mesh is skewed. In two dimensions, a minimum of two neighboring elements is necessary to solve the system. When the number of available neighbors is greater than two, then the system is over-determined. A classical way to solve this over-determined system is to multiply both sides of $AX = B$ by the transpose matrix. A square system is obtained: $A^T AX = A^T B$, and the solution reads, $X = (A^T A)^{-1} A^T B$.

It is possible to consider direct neighbors only of the considered cell (direct stencil) or both direct and indirect neighbors (extended stencil). Both configurations are schematized in Figure 8.1.

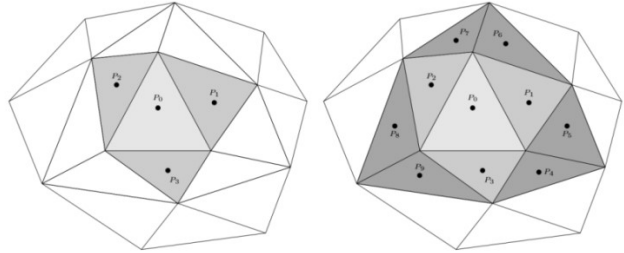

Figure 8.1: Schematic representation of the direct and indirect neighbors of the considered cell on an unstructured mesh made of triangles. On the left, only the direct neighbors are colored (direct stencil). On the right, both direct and indirect neighbors are colored (extended stencil).

In the following, extended stencil is retained for accuracy reasons. The liquid volume fraction gradient being determined, the next step consists in computing the interface curvature at each cell center.

### 8.2 Interface curvature determination at cell centers
The local curvature has been defined in (8.1) and requires volume fraction gradients computation, as detailed earlier. However, as the liquid volume fraction has been sharpened with the method given in Section 5 difficulties emerge. The liquid volume fraction gradient is not defined in a sufficiently wide stencil to compute curvature accurately.

To circumvent this difficulty, a color function $C_{liq}$ with smooth profile is introduced. It is built at each time step with the following initial data,

$$\begin{cases} \text{If } \alpha_{liq} \geq 0.5, \; C_{liq} = 1 \\ \text{Otherwise}, \; C_{liq} = 0 \end{cases}$$

An iterative type diffusion operator is then used as,

$$C_{liq,i}^{n+1} = \frac{S_i C_{liq,i}^{n} + \sum_{k=1}^{N_{Neigh}} S_k C_{liq,k}^{n}}{S_i + \sum_{k=1}^{N_{Neigh}} S_k},$$

with $N_{Neigh}$ the number of direct neighbors of the cell i and $S_i$ the surface of the cell i.

Using this operator during 5 iterations results in a diffuse profile on approximatively 5 cells. The corresponding color function is shown in Figure 8.2 in the specific case of a 2D liquid drop.

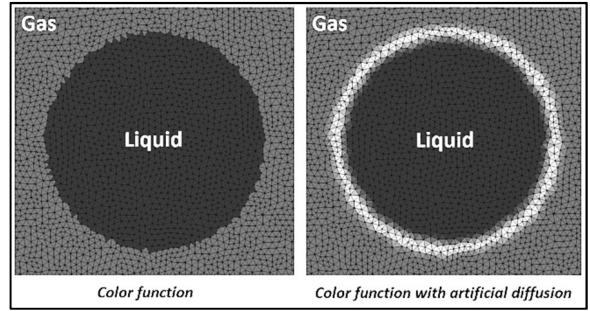

Figure 8.2: Color function representing a liquid drop surrounded by air on an unstructured mesh.

The diffused color function is therefore a good candidate for curvature calculation. The latter is then computed as:

$$\kappa = -\text{div}\left(\frac{\overrightarrow{\nabla C_{liq}}}{\left|\overrightarrow{\nabla C_{liq}}\right|}\right).$$

In two-dimension the local interface curvature at a cell center i reads,

$$\kappa_i = -\frac{\left(\frac{\partial C_{liq,i}}{\partial x}\right)^2 \frac{\partial^2 C_{liq,i}}{\partial y^2} - 2 \frac{\partial C_{liq,i}}{\partial x} \frac{\partial C_{liq,i}}{\partial y} \frac{\partial^2 C_{liq,i}}{\partial x \partial y} + \left(\frac{\partial C_{liq,i}}{\partial y}\right)^2 \frac{\partial^2 C_{liq,i}}{\partial x^2}}{\left|\overrightarrow{\nabla C_{liq,i}}\right|^3}. \tag{8.3}$$

This expression involves color function gradient components at cell centers, as well as color function Hessian matrix components. Computation of these terms is achieved with the help of the least-squares method detailed in Section 8.1 and used twice. First with the cell center color function as argument and second with resulting gradient components that become arguments of the Hessian matrix approximation.

**Curvature computation oscillations**

To validate the curvature computation method, a test with a simple circular interface is considered. By definition it is the inverse of the radius. Let us consider a 5mm radius sodium drop with curvature $200\,\text{m}^{-1}$.

The curvature obtained using (8.3) is shown in Figure 8.3, for two types of meshes (triangular cells and square ones). Whatever the mesh used, the computed curvature oscillates along the interface. It is thus necessary to correct its computation. This issue has been reported many times (Renardy and Renardy 2002, Garrick et al. 2017)

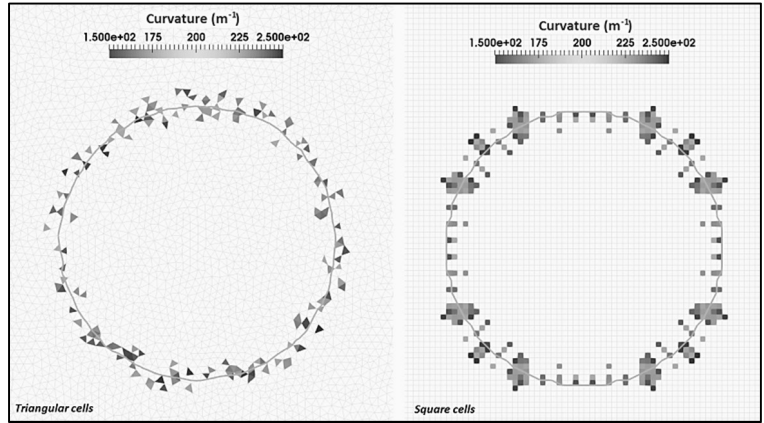

Figure 8.3: Computed curvature for a circular interface – Two types of meshes are considered – Only cells with curvature close to the theoretical value are shown. Large variations are present.

**Curvature correction**

Curvature computation is accurate at cells located on the interface, i.e. at cells for which the color function is close to 0.5. The method adopted consists in extending the value of the curvature computed in these cells to the surrounding ones, with the help of a weighted diffusion operator (Garrick et al., 2017). This is done with the following iterative method, used at each time step,

$$\kappa_i^{n+1} = \frac{\omega_i \kappa_i^n + \sum_{k=1}^{N_{Neigh}} \omega_k \kappa_k^n}{\omega_i + \sum_{k=1}^{N_{Neigh}} \omega_k},$$

with $\omega_k = S_k \left[ C_{liq,k} \left( 1 - C_{liq,k} \right) \right]^2$ a Gaussian weighting function centered on the interface where $C_{liq,k} \approx 0.5$. This correction is illustrated in Figure 8.4 with the same 5 mm radius sodium drop test case as before.

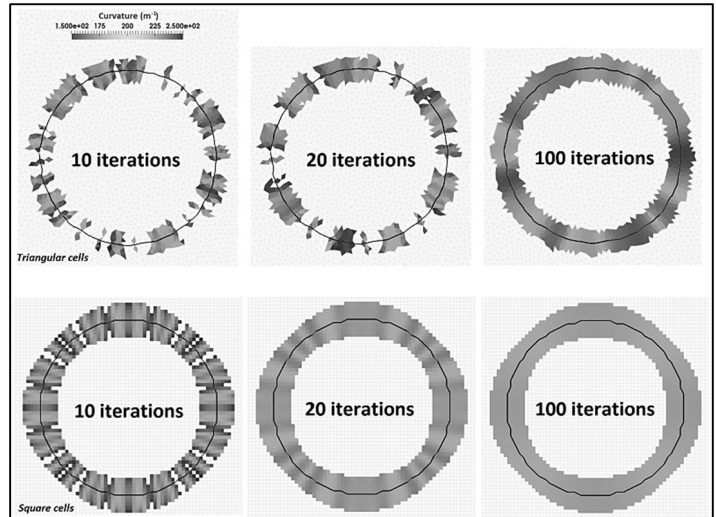

Figure 8.4: The circle curvature is computed with the two different types of meshes (triangles on top, squares on bottom). For each mesh type, varying numbers of iterations in the curvature diffusion method are used.

Accuracy of the curvature computation increases with the number of iterations of the diffusion method. However, we note that the convergence towards the theoretical value is faster in the case of square cells. However, it is important to limit corrections, especially in zones where the interface is highly curved, as neighboring cells values would corrupt the correct value of $\kappa$. In the various computations that will be presented in Section 9, Cartesian grids are used with typically 10 iterations. In the present application smooth interfaces are mainly considered and the method appears appropriate. To preserve mechanical equilibrium extra ingredient is needed such as specific Riemann solver, as examined hereafter.

## 8.3. Capillary HLLC Riemann solver

In order to avoid numerical errors due to operator splitting, surface tension has to be taken into account in the Riemann solver. The same arguments as those associated to gravity effects hold.

Building HLLC-type Riemann solver including capillary effects follows the same lines as the one with gravity. Let us consider the hyperbolic part of System (2.1), in the simplified context of a mixture made of a single liquid and a gas. The Rankine-Hugoniot relations (7.3) through the extreme waves of speeds $S_L$ and $S_R$ (estimated as done in Section 7) are used again. The same expressions of star pressures given by (7.4) are recovered. Indeed, surface tension has effects only at contact waves, not across acoustic ones.

In the presence of surface tension pressure equality through the contact wave ($p_L^* = p_R^*$) is replaced by the following condition (Perigaud and Saurel, 2005, Garrick et al., 2017):

$$p_R^* - p_L^* = \sigma \kappa \left( \alpha_{liq,R} - \alpha_{liq,L} \right),$$

with $\kappa = \frac{1}{2}(\kappa_L + \kappa_R)$.

As the curvature as been smoothed in a band surrounding the interface, the simple average above has been found appropriate to estimate the curvature at cell boundaries.

The contact relation is in agreement with the Laplace law. Typical examples and verifications are illustrated in Figure 8.5.

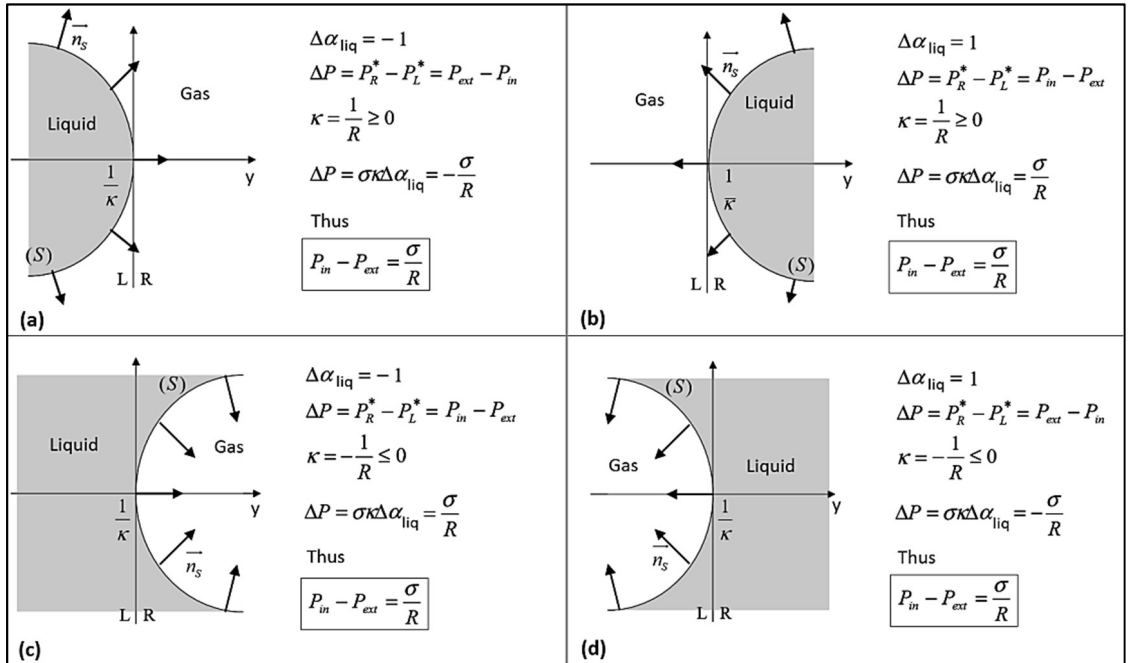

Figure 8.5: The various configurations that may occur at a curved interface separating a liquid and a gas.

The star pressure $p^*$ is then obtained as:

$$p^* = p_R^* - \sigma \kappa \alpha_{liq,R} = p_L^* - \sigma \kappa \alpha_{liq,L} \quad (8.4)$$

Combining (7.4) and (8.4) implies:

$$\begin{cases} S_M = \dfrac{p_R - p_L + \rho_L u_L (S_L - u_L) - \rho_R u_R (S_R - u_R) - \sigma \kappa (\alpha_{liq,R} - \alpha_{liq,L})}{\rho_L (S_L - u_L) - \rho_R (S_R - u_R)} \\ p^* = \dfrac{(p_R + \rho_R (u_R - S_R) u_R) \rho_L (u_L - S_L) - (p_L + \rho_L (u_L - S_L) u_L) \rho_R (u_R - S_R)}{\rho_L (u_L - S_L) - \rho_R (u_R - S_R)} - \sigma \kappa \dfrac{\alpha_{liq,R} \rho_L (u_L - S_L) - \alpha_{liq,L} \rho_R (u_R - S_R)}{\rho_L (u_L - S_L) - \rho_R (u_R - S_R)} \end{cases} \quad (8.5)$$

The first relation of System (8.5) corresponds to the one given in Garrick et al. (2017). The primitive variable vector in the star states is thus fully determined and the solution sampling is done to compute the flux F.

For example, in the subsonic case such as $S_M > 0$, the flux solution of the Riemann problem on a given face f along the face normal vector $\vec{\eta_f}$ reads,

$$\left(F.\vec{\eta_f}\right)^* = \begin{pmatrix} \rho_L^* Y_{k,L}^* S_M \\ \rho_L^* S_M \vec{u_L^*} + p_L^* \vec{\eta_f} \\ \left(\rho_L^* E_L^* + p_L^*\right) S_M \end{pmatrix}, \text{ with } \begin{cases} Y_{k,L}^* = Y_{k,L} \\ \rho_L^* = \rho_L \dfrac{u_L - S_L}{S_M - S_L} \\ E_L^* = E_L + \dfrac{p_L u_L - p_L^* S_M}{\rho_L (u_L - S_L)} \\ \vec{u_L^*} = S_M \vec{\eta_f} + \vec{u_L^{\perp}} \end{cases}.$$

In these formulas, $S_M$ is defined by (8.5) and $p_L^*$ is given by (7.4) or alternatively by (8.4) combined with (8.5). Contrary to the gravity case, the pressure $p^*$ is not used in the flux F computation when capillary effects are considered. Indeed, presence of a pressure discontinuity at equilibrium state requires pressure computation from (7.4).

When both gravity and surface tension effects are present, the preceding estimate of $S_M$ is replaced by:

$$S_M = \frac{p_R - p_L + \rho_L u_L (S_L - u_L) - \rho_R u_R (S_R - u_R) + \rho_R g (y_R - \bar{y}) - \rho_L g (y_L - \bar{y}) - \sigma \kappa (\alpha_{liq,R} - \alpha_{liq,L})}{\rho_L (S_L - u_L) - \rho_R (S_R - u_R)} \quad (8.6)$$

In this context, the flux F computation in the subsonic case such as $S_M > 0$ would then become:

$$\left(F.\vec{\eta_f}\right)^* = \begin{pmatrix} \rho_L^* Y_{k,L}^* S_M \\ \rho_L^* S_M \vec{u_L^*} + \left[p_L^* + \rho_L g (y_L - \bar{y})\right] \vec{\eta_f} \\ \left[\rho_L^* E_L^* + p_L^* + \rho_L g (y_L - \bar{y})\right] S_M \end{pmatrix}, \text{ with } \begin{cases} Y_{k,L}^* = Y_{k,L} \\ \rho_L^* = \rho_L \dfrac{u_L - S_L}{S_M - S_L} \\ E_L^* = E_L + \dfrac{p_L u_L - \left[p_L^* + \rho_L g (y_L - \bar{y})\right] S_M}{\rho_L (u_L - S_L)} \\ \vec{u_L^*} = S_M \vec{\eta_f} + \vec{u_L^{\perp}} \end{cases}$$

In these formulas, $p_L^*$ is given by (7.4) and $S_M$ is defined by (8.6).

**8.4. Validation**

A 5mm radius sodium drop is placed on liquid water surface. The rest of the domain is made of ambient air. The various fluids are transported at velocity 0.5m/s to the right. Only hydrodynamic and surface tension effects of sodium are considered in these computations. The pressure jump, initially imposed inside the drop, is in agreement with the Laplace law:

$$p_{N_a} = p_{atm} + \frac{\sigma}{R} = 10^5 + \frac{1.5}{5 \times 10^{-3}} = 100300 Pa.$$

Here, $\sigma = 1.5 N.m^{-1}$ corresponds to a large value, in order to assess robustness of the numerical method. The aim of these computations (Figure 8.6) is to check pressure jump preservation during drop motion.

With the various ingredients it is now possible to address 2D computations.

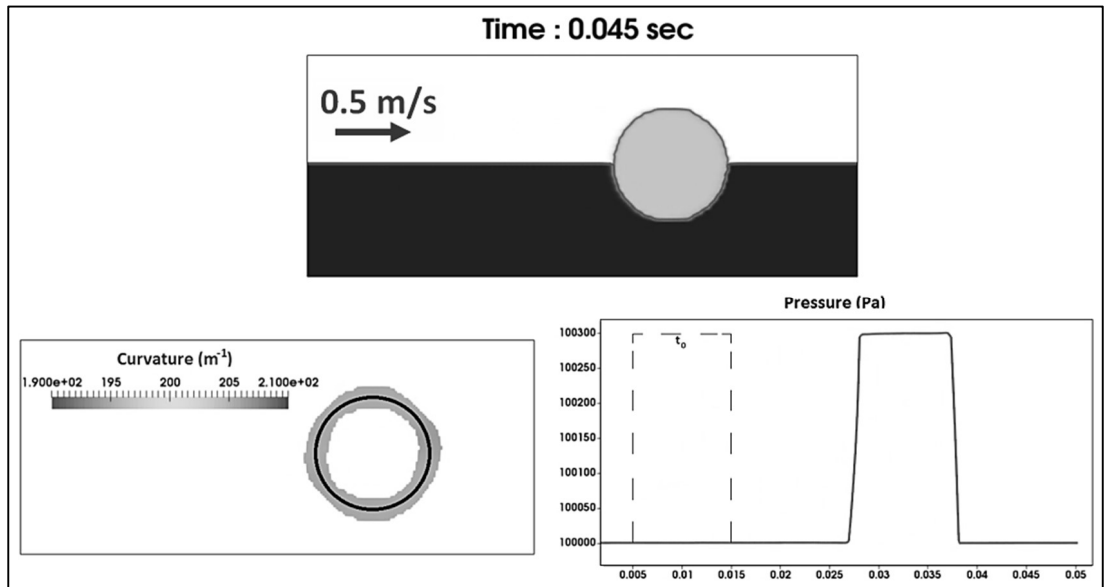
Figure 8.6: The initial pressure jump is preserved, validating the capillary solver.

## 9. 2D computation of the sodium/water flow with reactions
The various physical effects considered in System (2.1) are now considered to compute various 2D configurations typical of sodium/water flow configurations with surface and gas phase chemical reactions. Two 2D tests are considered.

**'Cold' liquid sodium drop (T=500K) in hot liquid water (T=370K)**
The initial configuration is shown in Figure 9.1.

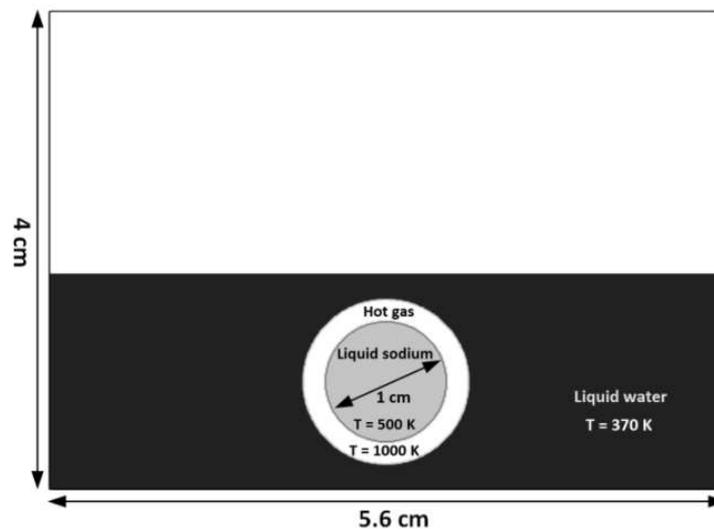
Figure 9.1: A liquid sodium drop is immersed in hot liquid water. Initially a hot gas (T=1000K) separates both liquids. Air is above liquid water surface. The entire domain is closed, and the pressure is initially atmospheric.

Computed results are shown in Figure 9.2. Four stages appear during time evolution.
a) During the first moments, the immersed liquid sodium drop reacts with the water vapor, through a diffusion flame. Pressure and temperature increase setting to motion the various liquid-gas interfaces and increasing the gas pocket volume.
b) The hot gas pocket reaches the liquid water surface and induces liquid film breakup. Surface waves propagate to the left and right of the liquid-gas interface while a liquid water fragment is projected vertically. The sodium drop moves up and the diffusion flame between sodium and water vapors increases intensity.

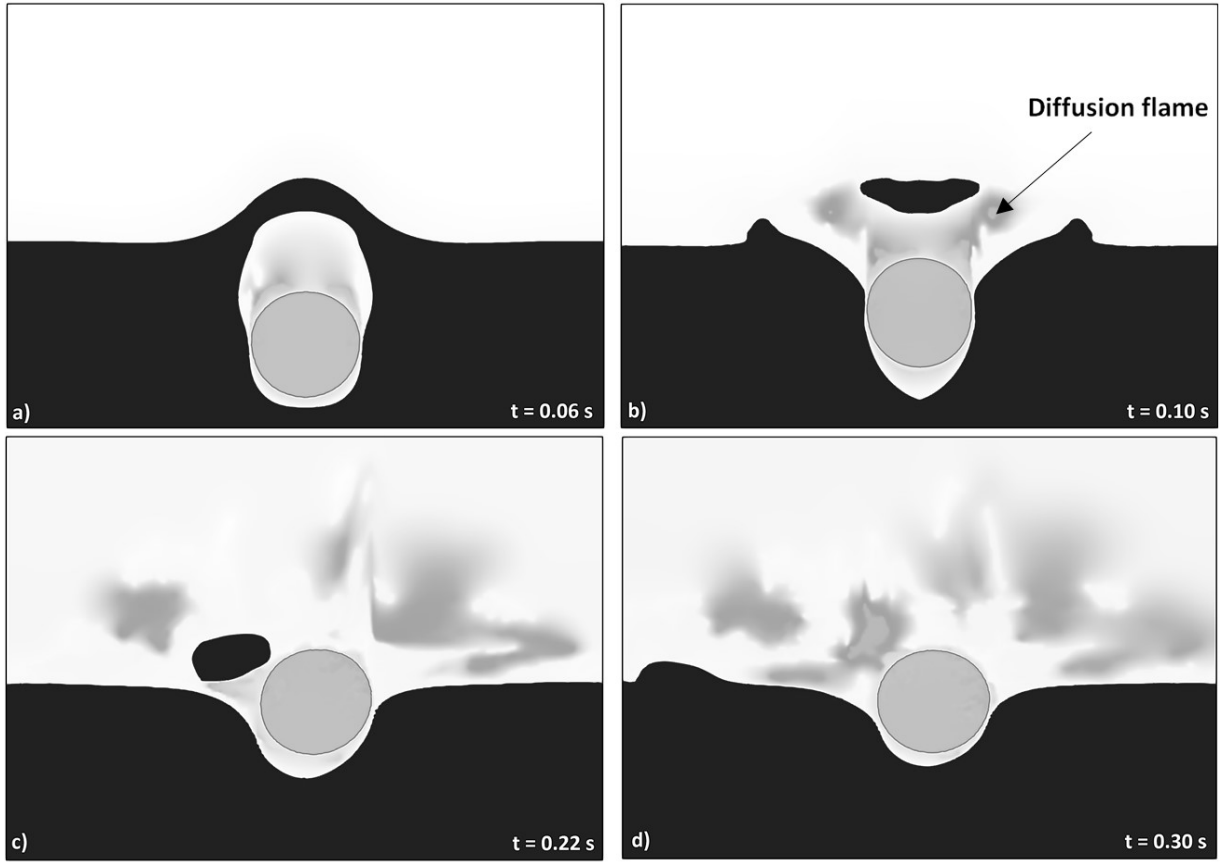

Figure 9.2: 2D computed results related to the test problem shown of Figure 9.1 in the presence of fluid motion, heat diffusion, phase transition, mass diffusion, both surface and gas reactions, gravity and capillary effects. The 2D mesh is made of 224 000 square cells. Results are shown at several times. The liquid water is shown in black through its volume fraction, whereas the liquid sodium drop is shown in grey. Elevated temperature zones are shown in shades of grey around the drop, showing the diffusion flame presence and burnt gases. Thermochemical Leidenfrost-type effect is reproduced.

c) The sodium drop is now in quasi-steady equilibrium at floats on the liquid water surface, still separated by the gas layer, whose thickness has been selected by the various diffusive effects, exothermicity of the reaction, weight of the sodium drop and gas flow in the interstice. The drop continues its autonomous motion as reported in the experiments.

The Leidenfrost-type effect, due to thermochemical transformations is thus at least qualitatively reproduced with the flow model and numerical method developed in the former sections. In the authors knowledge this is the first time such autonomous motion is reproduced numerically.

According to SWR experiments made at CEA Cadarache (Daudin, 2015), explosions seem to appear when sodium vaporization becomes significant. The boiling temperature at atmospheric pressure is about 1156K while in the present computations, the liquid sodium temperature is about 530 K at the end of the computations, corresponding to a significant temperature rise of 30K in 0.3s. Very long computations would be needed to reach the boiling temperature and we prefer to start from different initial conditions, with a hot liquid sodium drop.

**Hot liquid sodium drop (T=1140K) in hot liquid water (T=370K)**
The initial configuration is schematized in Figure 9.3. Computed results are shown in Figure 9.4. The initial temperature of the liquid sodium drop (1140K) is close to the boiling one at atmospheric pressure ($\approx$ 1156K). In this context, a large amount of vapor sodium appears quickly because of water vaporization, mass diffusion through the gas layer and surface reaction. Reaction follows in the gas layer that becomes quickly a large diffusion flame. Both temperature and pressure rise, increasing gas volume and setting intense fluid motion. These events are shown on graph (a). Phase change becomes intense in the sodium drop until the boiling point is reached. A large amount of sodium vapor continues to react with the water vapor. The flame exits the computational domain, with open outflow (b).

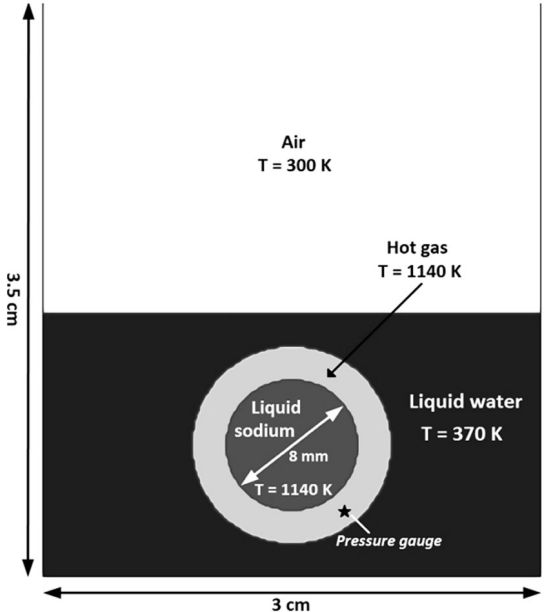

Figure 9.3: A hot liquid sodium drop is immersed in hot liquid water. Initially a hot gas (T=1140K) separates both liquids. Air is above the water. The pressure is initially atmospheric in the whole domain. The top of the domain is treated as an outflow/inflow boundary condition as detailed in Appendix C, the tank being made of air at atmospheric conditions. A static pressure gauge is placed within the initial hot gas layer and is shown by a star symbol.

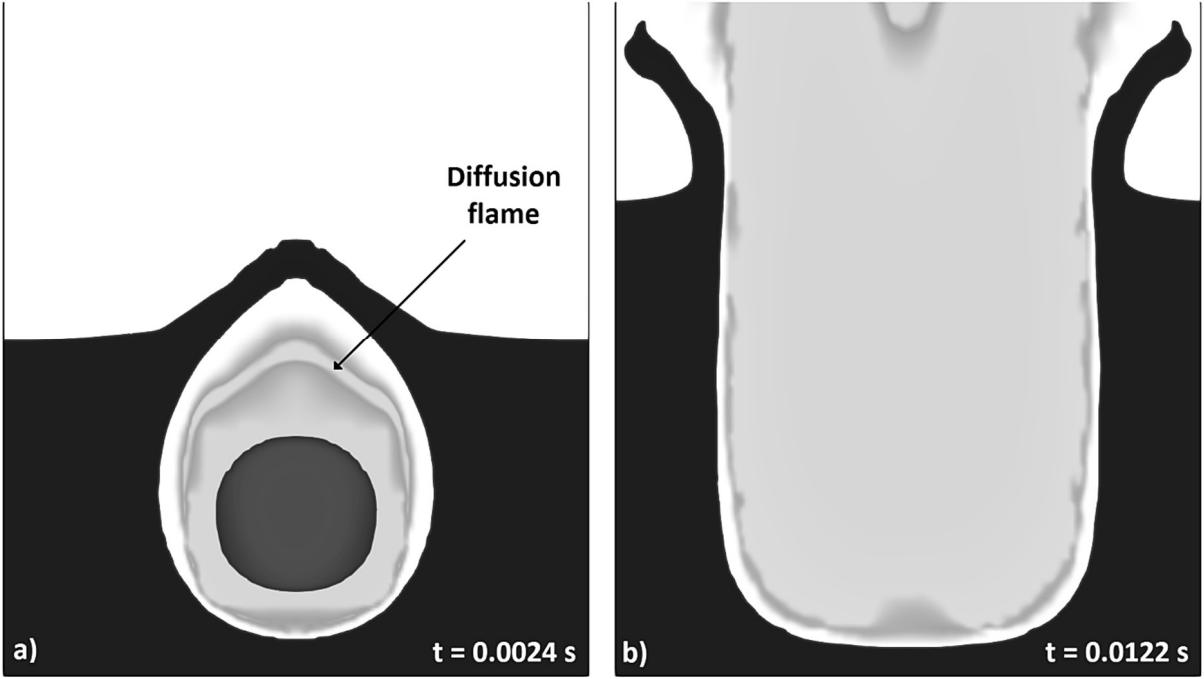

Figure 9.4: 2D computed results related to the 2D test problem of Figure 9.3 with both hot sodium and hot water. These computations consider compressible fluid motion, heat diffusion, phase transition, mass diffusion, both surface and gas reactions, gravity and capillary effects. The 2D mesh is made of 51 450 square cells. Results are shown at several times. The liquid water is shown in black through its volume fraction, whereas the liquid sodium drop appears in dark grey. Elevated temperature zones are shown in shades of grey around the drop, denoting the diffusion flame presence and burnt gases.

With the present initial data, the diffusion flame is much more intense than previously, as expected. But the reaction energy is released gradually, as for any diffusion flame, forbidding appearance of explosion effect with shock wave emission. The pressure recorded by the sensor and the graph representing the maximum temperature reached in the gas phase versus time confirm the previous observations (Figure 9.5).

As shock waves are observed in the experiments extra physical effects must be considered, meaning that flow model (2.1) is not enough representative.

The diffusion flame temperature (>3000K) observed in the previous computations is significantly higher than the one measured in SOCRATE experiments (Carnevali, 2012), where flame temperature of the order of 1670K are reported. Until now, soda phase transition has not been considered. As vaporization of soda is highly endothermic, it is very likely that the flame temperature will become more realistic if the phase transition of soda is considered.

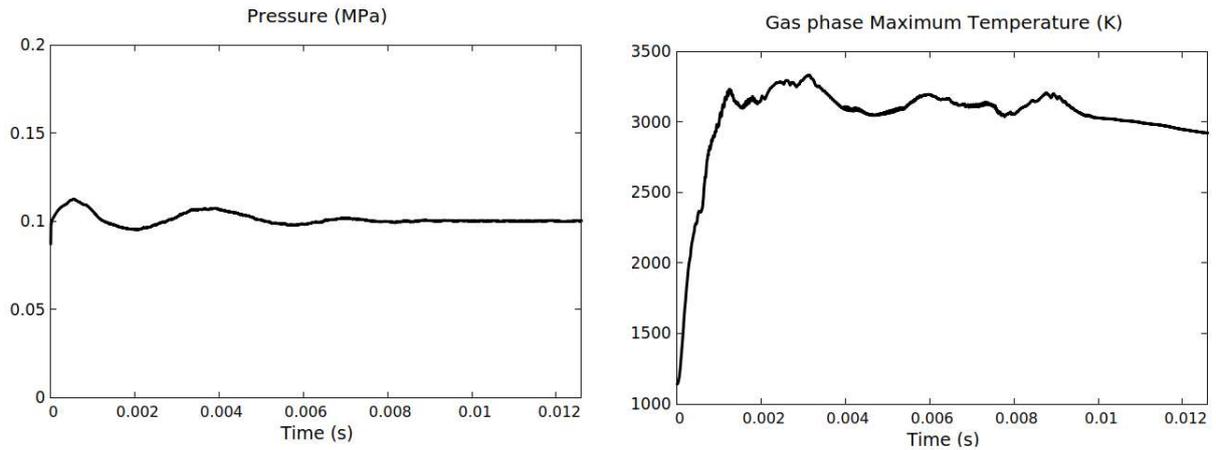

Figure 9.5: On the left, the pressure signal recorded by the sensor versus time confirms absence of shock waves. The sensor location is specified in Figure 9.3. During the whole simulation, the sensor is located within the gas film separating sodium and water liquids. On the right, the maximum temperature reached in the gas phase versus time is shown. This one exceeds 3000K and illustrates the flame intensity.

**Soda phase transition**

Reference data such as latent heat of vaporization and saturation pressure are reported in (Lide, 2009) and have been used to determine soda vapor NASG parameters, knowing those of liquid soda (see Table 3.1).

The mass balance equation for soda vapor is thus added to System (2.1):

$$\frac{\partial \rho Y^g_{Na_aOH}}{\partial t} + \text{div}\left(\rho Y^g_{Na_aOH} \vec{u} + \alpha_g \vec{F}^g_{Na_aOH}\right) = -\rho \nu_{Na_aOH} \left(g^g_{Na_aOH} - g^L_{Na_aOH}\right) \tag{9.1}$$

The mass diffusion flux $\vec{F}^g_{Na_aOH}$ ensures the molecular diffusion of the soda vapor within the gas film. Its modelling is identical to the one used for the other gaseous species (Appendix A). Liquid-vapor mass transfer is treated through Gibbs free energy relaxation, in the same way as for water and sodium. Its complementary source term is introduced in the liquid soda mass balance equation as,

$$\frac{\partial \rho Y^L_{Na_aOH}}{\partial t} + \text{div}\left(\rho Y^L_{Na_aOH} \vec{u}\right) = \rho \nu_{Na_aOH} \left(g^g_{Na_aOH} - g^L_{Na_aOH}\right). \tag{9.2}$$

Local thermodynamic equilibrium $\left(\nu_{Na_aOH} \to \infty\right)$ is again assumed, meaning that the thermochemical solver developed in Chiapolino et al (2017) (Appendix B) is used.

As soda is treated more physically, its presence and effects are reconsidered. Despite the consideration of soda phase transition, liquid soda at high temperature is still expected to be present. Liquid soda being heavier than liquid water, it has the capability to precipitate in water. Such a hot liquid penetrating water may have serious consequences on water vapor production. This option is examined in the following as a possible source of explosion.

**Liquid soda diffusion and mixing with liquid water**

The diffuse interface model considered presents a nice property in the sense that it is able to create interfaces not present initially. Indeed, liquid soda is absent in the initial conditions and appears dynamically in the computations as a liquid layer. At this level it is difficult to figure out the importance of surface tension effects for this soda layer. They may result in soda liquid droplet appearance, whose penetration under water seems possible. But knowledge and observations at this level are rare and a simplified approach is preferred for a first qualitative study of soda effect. Penetration of liquid soda in liquid water is considered through a diffusion model, for simplicity reasons.

In the gas phase, the various species are ideally mixed and mass diffusion occurs through the approach detailed in Appendix A. In this frame each gas species occupies the entire gas volume and has its own partial pressure, as well known with the Dalton's law. Different modelling is needed to address diffusion of liquids. Indeed, thermodynamics of a given liquid phase is valid only in its own volume. As the liquid mixture and gas mixture evolve in both pressure and temperature equilibria the Gibbs identity of a given liquid phase reads,

$$Tds_k = de_k + pdv_k.$$ (9.3)

which is significantly different of the one used in Appendix A for gas mixtures,

$$Tds_k = de_k + p_k dv_k,$$

where $p_k$ denotes the partial pressure of a given chemical species in the gas mixture.

Mass diffusion modelling within the liquid phase must fulfill the second law of thermodynamics. Reconsidering the derivation done in Appendix A with (9.3) yields to the following mass diffusion flux,

$$F_{NaOH} = -\frac{\partial p}{\partial x}\left(v^L_{NaOH} - v^L_{H_2O}\right)c^{H_2O}_{NaOH},$$ (9.4)

with $c^{H_2O}_{NaOH}$ the diffusion coefficient of the species $N_aOH$ into the species $H_2O$ in the liquid phase.

The associated mixture entropy inequality, in the same simplified calculation frame as in Appendix A yields now,

$$\frac{\partial \rho s}{\partial t} + \frac{\partial\left(\rho su + (s_{NaOH} - s_{H_2O})F_{NaOH}\right)}{\partial x} = \frac{1}{2T}\left(\frac{\partial p}{\partial x}\right)^2 \left(v^L_{NaOH} - v^L_{H_2O}\right)^2 c^{H_2O}_{NaOH} > 0$$

It should be noted that the diffusion law (9.4) is reminiscent of Darcy's law. Contrarily to mass diffusion in gas mixtures that is mainly driven by concentration gradients, here mass diffusion is driven by mixture pressure gradient. In multi-D configurations and in the present sodium-water flow context, pressure gradient is mainly hydrostatic ($\vec{\nabla}p = \rho\vec{g}$). The present mass diffusion modelling enables liquid soda precipitation in liquid water.

In the following computations the mass diffusion coefficient is taken constant and equal to $10^{-4}\,kg.s/m^3$. Influence of this coefficient is not particularly examined in the following as expected effects and particularly explosions are not observed. To be more precise the 2D test problem of Figure 9.3 with both hot sodium and hot water is reconsidered. Soda phase transition, liquid soda mass diffusion as well as the various physical effects considered in System (2.1) are activated.

Computed results are shown in Figure 9.6. Qualitatively, the results are similar to those of Figure 9.4, apart the enhanced flame thickness due to the molecular diffusion of soda in vapor state. As before, a very intense diffusion flame is generated by the reaction of water and sodium vapors. Mass diffusion of hot liquid soda in liquid water just results in enhanced water vapor production, rendering the diffusion flame more intense. But shock waves are still absent as shown in Figure 9.7. The only noticeable fact with respect to experimental observations is the computed flame temperature (about 1600K) which becomes much more realistic thanks to the consideration of soda phase transition.

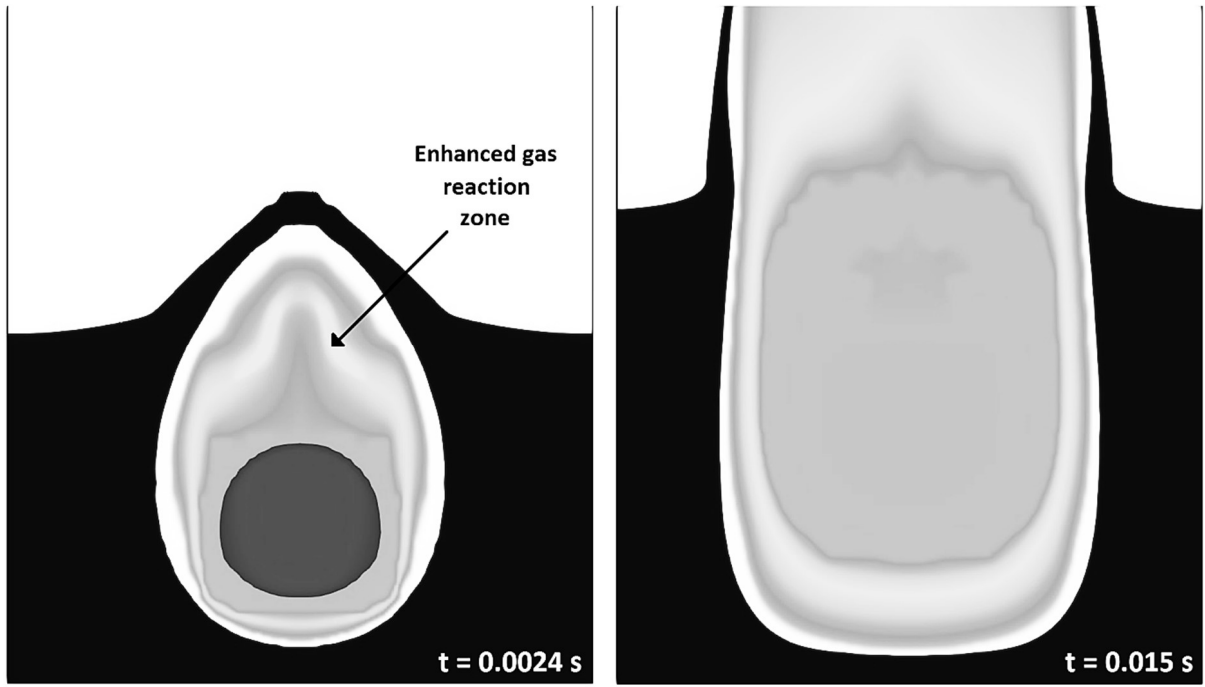

Figure 9.6: 2D computed results related to the 2D test problem of Figure 9.3 in the presence of compressible fluid motion, heat diffusion, phase transition of sodium, water and soda species, mass diffusion in the gas phase, liquid soda mass diffusion, surface and gas reactions, gravity and capillary effects. The 2D mesh is made of 51 450 square cells. Results are shown at two times. The liquid water is shown in black through its volume fraction, whereas the liquid sodium drop appears in dark grey. Elevated temperature zones are shown in shades of grey around the drop.

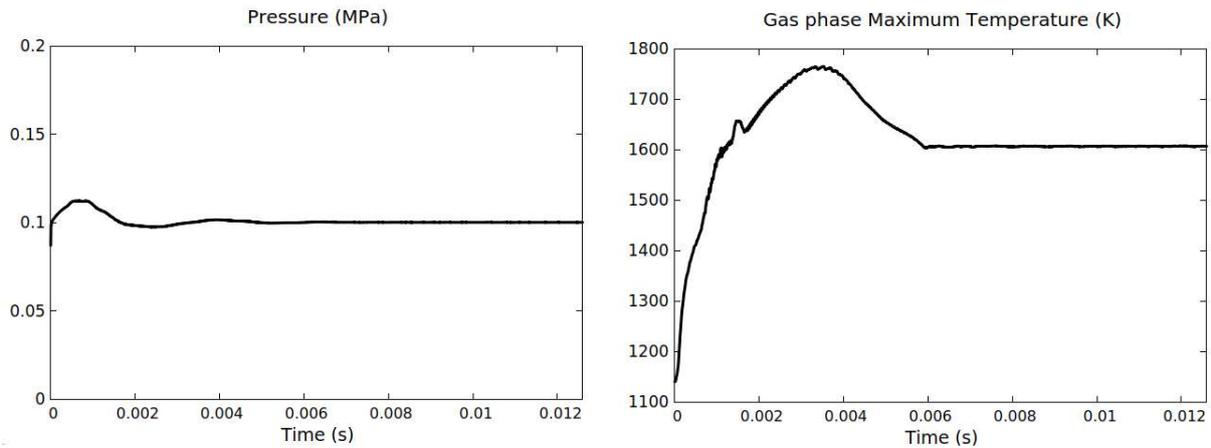

Figure 9.7: On the left, the pressure signal recorded by the sensor versus time confirms shock waves absence, despite the consideration of liquid soda mass diffusion. On the right, the maximum temperature of the gas phase relaxes to the experimental one reported in (Carnevali, 2012).

Extra computations have been carried out with initially colder sodium drops. Liquid soda mass diffusion resulted in enhanced water vapor production, resulting in enhanced sodium drop surface reaction and significant reduction of sodium heating time. However, it didn't lead to shock wave emission as the flame remained a diffusion one. To progress in the understanding of the mechanism responsible for shock wave appearance, other options are considered in the following.

Up to now the gas phase reaction has been considered as instantaneous resulting in the impossibility of reactant gas premixed zones appearance. Finite rate chemistry effects are thus addressed hereafter.

**Delayed ignition**

Chemical kinetics data for the reaction between water and sodium vapors seem absent in the literature. An estimation of the gas reaction rate is given in Takata et al. (2009) via the transition state theory, but in the author's knowledge, no experiments have been done to validate related data. Considering delayed gas reaction allows mixing of water and sodium vapors present in the gas film before reaction triggering. If gas premixing is intense enough, a premixed flame is expected instead of a diffusion one. The effects of such premixed flame are investigated in the following.

The gas reaction between water and sodium vapors is recalled hereafter:

$$Na^{(v)} + H_2O^{(v)} \rightarrow NaOH^{(L)} + \frac{1}{2}H_2^{(g)}.$$

Finite-rate chemistry is addressed through the following mass production rates,

$$\frac{\partial \rho Y_{Na}^g}{\partial t} = \dot{\omega}_{GR}^{Na} \qquad \frac{\partial \rho Y_{H_2O}^g}{\partial t} = \dot{\omega}_{GR}^{H_2O} \qquad \frac{\partial \rho Y_{NaOH}^L}{\partial t} = \dot{\omega}_{GR}^{NaOH} \qquad \frac{\partial \rho Y_{H_2}^g}{\partial t} = \dot{\omega}_{GR}^{H_2},$$

with

$$\begin{cases} \dot{\omega}_{GR}^{Na} = -W_{Na} k_f [Na][H_2O] \\ \dot{\omega}_{GR}^{H_2O} = -W_{H_2O} k_f [Na][H_2O] \\ \dot{\omega}_{GR}^{NaOH} = W_{NaOH} k_f [Na][H_2O] \\ \dot{\omega}_{GR}^{H_2} = \frac{1}{2} W_{H_2} k_f [Na][H_2O] \end{cases}$$

with $[k]$ the molar concentration of the species k in the gas phase.

An Arrhenius-type law is used for the reaction rate $k_f$,

$$k_f(T) = k_0 T^n e^{-E_a/RT}, \tag{9.4}$$

with $k_0$ the pre-exponential factor, n a real, $E_a$ the activation energy and $R = 8.314 \, J/mol/K$ the gas constant. Reaction rate constants have been chosen so that the activation energy is reached for temperatures of the order of 1250 K:

$k_0 = 9 \times 10^{20}$, $\qquad n = 2$, $\qquad E_a = 5.75 \times 10^5 \, J/mol$.

These constants result in stiff reaction when the threshold temperature is reached. This modeling is quite arbitrary but allows qualitative analysis of premixed flame effects. An extra effect is considered as well to enhance gas mixing.

**Turbulent diffusion**

Experiments carried out by Daudin (2015), Daudin et al. (2018) have highlighted the turbulent nature of the gaseous film. Many gas bubbles of various sizes separate the sodium drop from the liquid water surface (Figure 9.8) rendering the gas film highly turbulent.

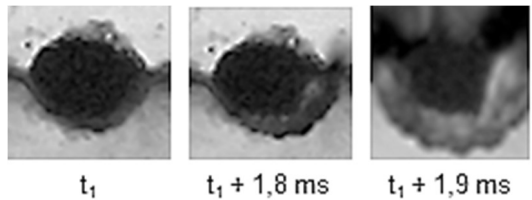

Figure 9.8: Gas bubbles present in the gas film separating the sodium drop and the liquid water surface (time t1 before explosion) - VIPERE experiment carried out at CEA Cadarache (Daudin 2015, Daudin et al. 2018). The gas film is consequently more a mixing zone than a film defined by two sharp interfaces.

Presence of these small bubbles are expected to enhance water and sodium vapors mixing prior to gas reaction runaway. Considering such turbulent mixing with liquid-gas interfaces could be addressed with the present flow model (System 2.1 and its extensions) but would require tremendous spatial and temporal resolution. For simplicity reasons, its effects are considered qualitatively through enhanced molecular diffusion within the gas film.

The mass diffusion coefficient C from the relationship (6.1) is modified as follows:

$$C\left(Y_{H_2O}^L, Y_{H_2O}^g, Y_{Na}^g\right) = C_0 H_\chi\left(Y_{H_2O}^L\right)\left(1 + a Y_{H_2O}^{g\,2} Y_{Na}^{g\,2}\right),$$

where $a = 3000$ represents an intensification parameter.

In the following 2D test based on the initial configuration of Figure 9.3, the various physical effects considered in System (2.1) are activated. The gas reaction with delayed ignition and turbulent diffusion are considered for this run, as well as soda phase transition and liquid soda mass diffusion previously introduced.

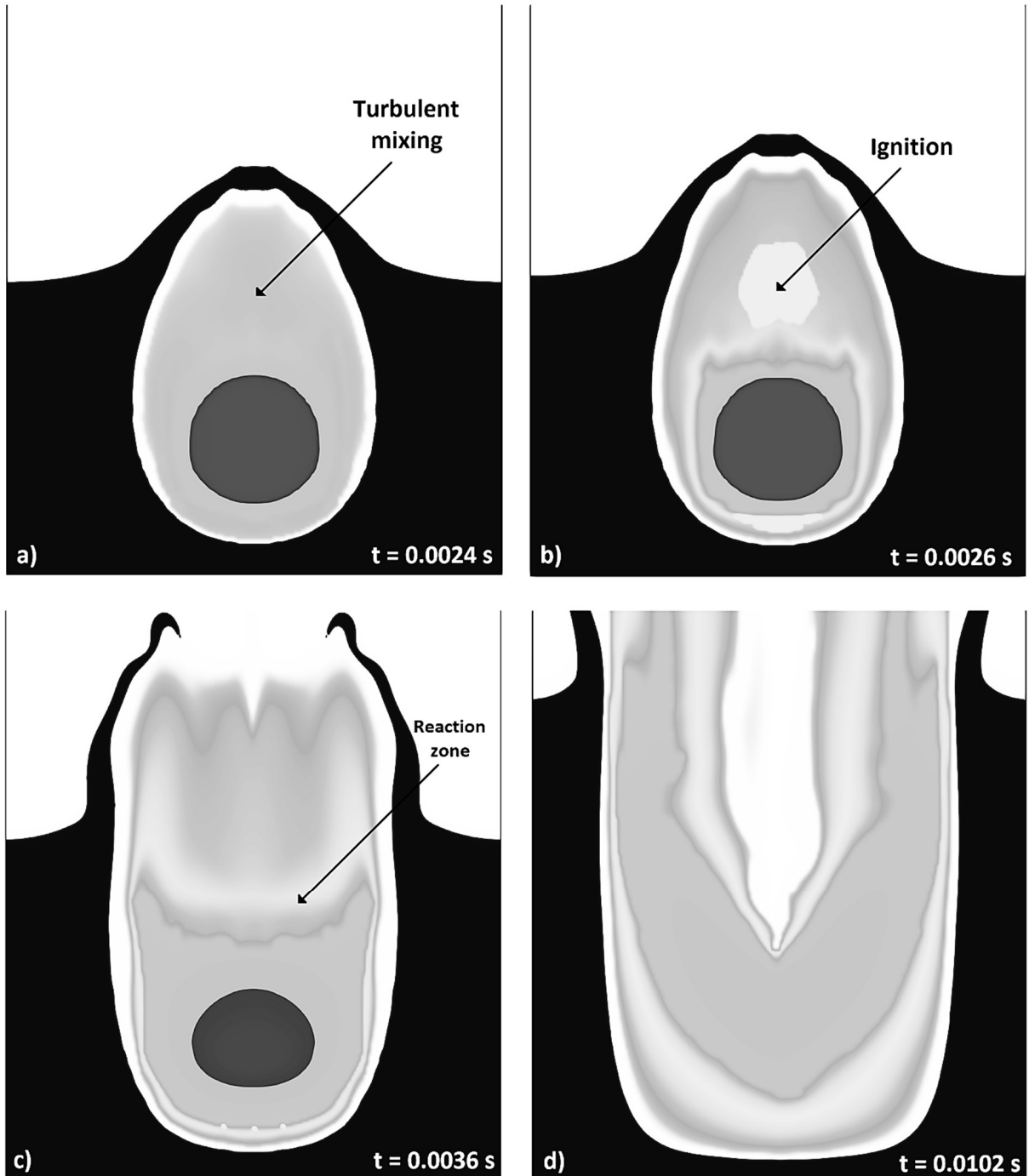

Figure 9.9: 2D computed results related to the 2D test problem of Figure 9.3 in the presence of compressible fluid motion, heat diffusion, phase transition of sodium, water and soda species, mass diffusion in the gas phase, liquid soda mass diffusion, surface reaction, gas reaction with delayed ignition, gravity and capillary effects. Turbulent mixing of sodium and water vapors has also been considered in the gas film. The 2D mesh is made of 51 450 square cells. Results are shown at several times. The liquid water is shown in black through its volume fraction, whereas the liquid sodium drop appears in dark grey. Elevated temperature zones are shown in shades of grey around the drop.

Computed results are shown in Figure 9.9. During the first instants, the immersed liquid sodium drop reacts with the water vapor through the surface reaction. Pressure and temperature increase setting to motion the various liquid-gas interfaces and increasing gas pocket volume. Sodium and water vapors present in large concentration in the gas film mix intensively, the activation energy of the gas reaction being unreached at this stage (a). When the gas pocket temperature locally reaches the threshold one (1250K) premixed vapors strongly react. A flame ignites and propagates in the gas layer (b), while sodium and water vapor are still produced at interfaces. Premixed flame propagation results in shock wave emission that deforms the liquid sodium drop and induces liquid water film breakup (c). The sodium boiling point has now been reached. The water and sodium vapors present in large amounts react and feed the flame whose shape has been disturbed by shock wave reflections (d).

The pressure recorded by the sensor and the graph representing the maximum temperature reached in the gas phase versus time are shown in Figure 9.10. The pressure graph shows the appearance of a shock wave at the time of ignition (t=0.0026s) corresponding to graph b) in Figure 9.9. The overpressure measured by the sensor in the gas pocket is around 0.06MPa. At the same time, the maximum temperature reached in the gas phase sharply increases. The flame temperature relaxes later around 1600K, as observed in the experiments.

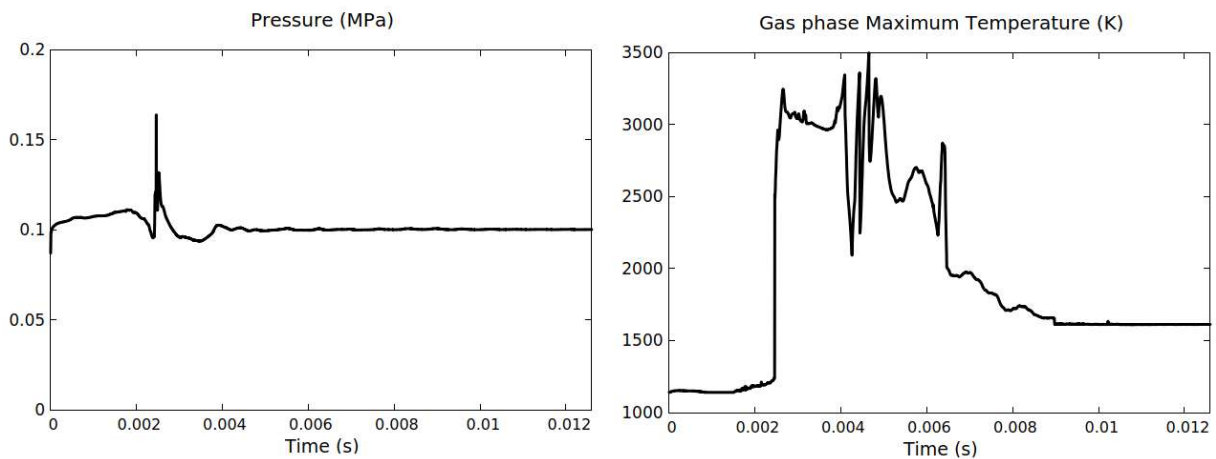

Figure 9.10: On the left, the pressure signal recorded by the sensor versus time shows the appearance of a shock once the activation energy of the gas reaction is reached, at time t=0.0026s. On the right, at the same time, the maximum temperature reached in the gas phase sharply increases. Later, the flame temperature relaxes to 1600K approximately.

The numerical results shown in Figure 9.9 can be qualitatively compared to photographs made in VIPERE experiments at CEA Cadarache (Daudin 2015), involving a sodium drop initially immersed in liquid water. At explosion time (Figure 9.11), the same type of flow behavior as the one reported in Figure (9.9) is observed.

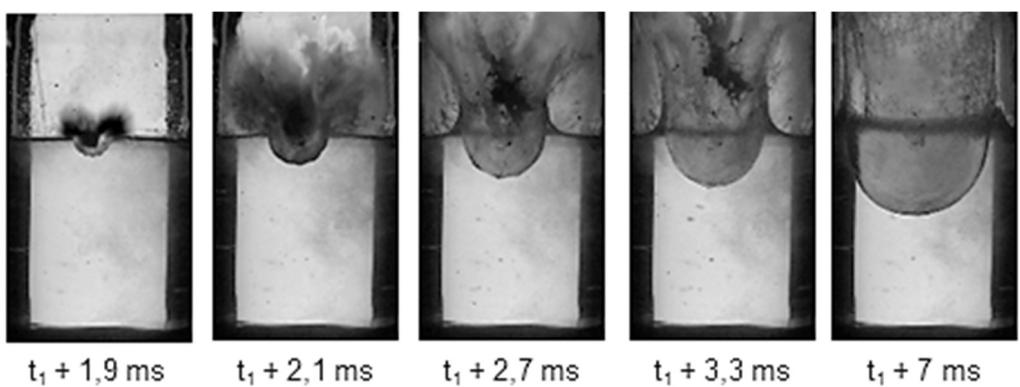

Figure 9.11: Photographs made in the VIPERE experiment carried out at CEA Cadarache (Daudin 2015) during explosion.

It therefore appears that explosion can be qualitatively reproduced by the present model and is a consequence of gas premixing and delayed ignition.

**10. Conclusion**

A diffuse interface method, consisting in a flow model in temperature and pressure equilibrium has been used as a basis to model contact and reaction between sodium and liquid water through a gas film. Many physical and chemical effects have been considered such as:
- sodium surface reaction,
- gas phase reaction,
- surface tension and buoyancy,
- phases compressibility,
- phase transition at interfaces,
- heat and mass diffusion,
- soda production and related phase transition,
- turbulent mixing in the gas layer.

Thanks to this modelling, two main effects have been qualitatively reproduced:
- The thermo-chemical Leidenfrost-type effect responsible for autonomous drop motion, as shown in Figure 9.2, in accordance with experimental observations;
- Explosion effect with shock wave emission (Figures 9.9 and 9.10).

It seems that the present work provides the first qualitative reproduction of these events with a numerical approach.

From the present computations, it seems that explosion is a result of turbulent mixing of the various vapors, resulting in a premixed flame and not a diffusion one, as speculated formerly.

Many perspectives emerge to improve present qualitative observations:
- Determination of the chemical kinetics of sodium vapor with water vapor;
- Use fine enough resolution to capture small scale physics such as bubbles dynamics in the vicinity of the gas film, responsible of turbulent mixing;
- Improve knowledge of mass diffusion coefficients.

A key point relies in the numerical treatment of surface tension effects. Two regularizations have been used in the present study, with the color function and curvature computation. Efforts must be done to improve related computations. Another numerical issue is related to the consideration of mass diffusion in the context of diffuse interface formulations. The difficulty has been outlined in Section 6 with a possible cure but needs extra efforts to become general.

**Appendix A. Modeling mass diffusion within the gas phase**

Insertion of mass diffusion effects in the gas phase in the context of the diffuse interface formulation is the aim of this appendix. The modelling of mass diffusion follows the lines of Giovangigli (2012) for gas mixtures with mild modifications as two phases are present in the present context. In System (2.1), the mass diffusion flux of a given gaseous species k is expressed as:

$$\vec{F}_k = C \frac{1}{p} \left( y_k \vec{\nabla} p - \vec{\nabla} p_k \right), \quad (A.1)$$

with C represents a diffusion coefficient (taken constant and equal to $10^{-4}\,\text{kg/m/s}$ in the present application). The associated energy flux reads:

$$\vec{q}_M = \sum_{k>3} h_k \vec{F}_k \,.$$

Compatibility of formulation (A.1) with the second law of thermodynamics is examined. Let us consider the simplified situation of a two-phase mixture made of liquid "L" and two-component gas "g", in 1D configuration. Mass diffusion only occurs within the gas mixture. Each phase (liquid and two-species gas) is assumed to occupy its own volume, and the gas mixture obeys Dalton's law.

To clarify the analysis, the model is written as follows, as a simplification of (2.1) in the absence of capillary effects, phase change and chemical reactions:

$$\begin{aligned}
\frac{\partial \alpha_g \rho_g y_1}{\partial t} + \frac{\partial \alpha_g \rho_g y_1 u + \alpha_g F_1}{\partial x} &= 0 \\
\frac{\partial \alpha_g \rho_g y_2}{\partial t} + \frac{\partial \alpha_g \rho_g y_2 u + \alpha_g F_2}{\partial x} &= 0 \\
\frac{\partial \alpha_L \rho_L}{\partial t} + \frac{\partial \alpha_L \rho_L u}{\partial x} &= 0 \\
\frac{\partial \rho u}{\partial t} + \frac{\partial \rho u^2 + p}{\partial x} &= 0 \\
\frac{\partial \rho E}{\partial t} + \frac{\partial u(\rho E + p) + \alpha_g Q}{\partial x} &= 0
\end{aligned} \quad (A.2)$$

$y_1$ and $y_2$ represent the mass concentrations within the gas phase ($y_1 + y_2 = 1$).

Mass concentrations of the phases may be used as well,

$$Y_g = \frac{\alpha_g \rho_g}{\rho} \text{ and } Y_L = \frac{\alpha_L \rho_L}{\rho}\,.$$

Terms $F_1$ and $F_2$ represent mass diffusion terms and $Q = \sum_k h_k F_k = (h_1 - h_2) F_1$ represents the associated energy flux. Expressions for $F_1$ and $F_2$ have to be determined. Obviously, $F_1 + F_2 = 0$.

The total energy is defined by,

$$E = Y_1 e_1 + Y_2 e_2 + Y_L e_L + \frac{1}{2} u^2,$$

with $Y_1 = y_1 Y_G$ et $Y_2 = y_2 Y_G$.

System (A.2) is expressed with Lagrangian derivatives:

$$\frac{dy_1}{dt} = -\frac{1}{\alpha_g \rho_g}\frac{\partial \alpha_g F_1}{\partial x}; \quad \frac{dy_2}{dt} = -\frac{1}{\alpha_g \rho_g}\frac{\partial \alpha_g F_2}{\partial x}; \quad \frac{dY_g}{dt} = 0; \quad \frac{dY_L}{dt} = 0; \quad \frac{d\rho}{dt} = -\rho\frac{\partial u}{\partial x};$$

$$\frac{du}{dt} = -\frac{1}{\rho}\frac{\partial p}{\partial x} \quad \text{and} \quad \frac{de}{dt} + p\frac{dv}{dt} + \frac{1}{\rho}\frac{\partial \alpha_g Q}{\partial x} = 0 \tag{A.3}$$

To obtain the entropy equation of the two-phase mixture the gas mixture one has to be determined first. It is based on the Gibbs identity of a given gas species,

$$T\frac{ds_k}{dt} = \frac{de_k}{dt} + p_k \frac{dv_k}{dt}.$$

Note that the partial pressure appears in this definition. In the specific context of ideal gas mixtures, specific volumes $v_k$ are defined as:

$$y_k v_k = v_g.$$

From the Gibbs identity of each gas component the gas mixture Gibbs identity is obtained as,

$$T\frac{d(y_1 s_1 + y_2 s_2)}{dt} = \frac{d(y_1 e_1 + y_2 e_2)}{dt} + p_1 \frac{dy_1 v_1}{dt} + p_2 \frac{dy_2 v_2}{dt} + (g_2 - g_1)\frac{dy_1}{dt}$$

with,

$$g_k = e_k + p_k v_k - T s_k.$$

We thus have,

$$T\frac{ds_g}{dt} = \frac{de_g}{dt} + p\frac{dv_g}{dt} + (g_2 - g_1)\frac{dy_1}{dt} \tag{A.4}$$

For the liquid phase, the Gibbs identity reads:

$$T\frac{ds_L}{dt} = \frac{de_L}{dt} + p\frac{dv_L}{dt} \tag{A.5}$$

with $v_L = \frac{1}{\rho_L}$.

Note that in (A.4) and (A.5) the pressure p is present instead of the partial one.

The two-phase mixture Gibbs identity is thus obtained as,

$$T\frac{ds}{dt} = \frac{de}{dt} + p\frac{dv}{dt} + (g_2 - g_1)\frac{dY_g y_1}{dt}, \tag{A.6}$$

where the mixture entropy is defined as $s = Y_L s_L + Y_g s_g$.

This result is reported in the energy equation (A.3):

$$T\frac{ds}{dt} + (g_1 - g_2)\frac{dY_g y_1}{dt} + \frac{1}{\rho}\frac{\partial \alpha_g Q}{\partial x} = 0$$

After some manipulations, we have:

$$\frac{\partial \rho s}{\partial t} + \frac{\partial \rho s u + (s_1 - s_2)\alpha_g F_1}{\partial x} = -\frac{\alpha_g F_1}{T}(v_1 \frac{\partial p_1}{\partial x} - v_2 \frac{\partial p_2}{\partial x})$$

With the mass diffusion fluxes defined by (A.1) non-negative entropy production is guaranteed:

$$\frac{\partial \rho s}{\partial t} + \frac{\partial \rho s u + (s_1 - s_2)\alpha_g F_1}{\partial x} = \frac{\alpha_g F_1^2}{T}\frac{p}{C}(v_1 + v_2) > 0$$

The molecular diffusion flux (A.1) can also be expressed as,

$$\vec{F}_k = C\frac{1}{p}\left((y_k - x_k)\vec{\nabla}p - p\vec{\nabla}x_k\right),$$

where $x_k$ denotes the molar fraction of the gas species k. In the specific context of the SWR most of the evolution is quasi-isobaric and the term $\vec{\nabla}x_k$ is the driving diffusion force.

## Appendix B. Stiff phase transition solver

Phase transition between a liquid and its vapor is a phenomenon that tends to relax the Gibbs free energy of the phases towards equilibrium. In System (2.1) related terms appear in the right-hand side of the mass balance equations, in the source term vector S summarized hereafter,

$$S = \left( \rho \nu_{H_2O} \left( g^g_{H_2O} - g^L_{H_2O} \right) \quad \rho \nu_{Na} \left( g^g_{Na} - g^L_{Na} \right) \quad 0 \quad -\rho \nu_{H_2O} \left( g^g_{H_2O} - g^L_{H_2O} \right) \quad -\rho \nu_{Na} \left( g^g_{Na} - g^L_{Na} \right) \quad 0 \quad 0 \quad 0 \right)^T.$$

where $g_k$ denotes the Gibbs free energy of the phase k defined in Section 2.

For a given species i (sodium or water), the parameter $\nu_i$ controls the rate at which thermodynamic equilibrium is reached. In the present context of SWR phase change is controlled by heat and mass diffusion, that are orders of magnitude slower than thermochemical relaxation. It is therefore appropriate and convenient to consider that $\nu_i$ tends towards infinity, meaning that local thermodynamic equilibrium is considered.

Chiapolino et al. (2017) have developed specific thermochemical solvers to determine the liquid-vapor mixture composition in the presence of a multicomponent gas mixture. This solver is fast, robust and accurate. In the SWR context, it is used to compute phase transition of water and sodium to their respective vapors. Each liquid-vapor phase transition is treated separately.

Let us consider phase transition of a given liquid with its vapor. Local thermodynamic equilibrium implies,

$$g_{vap}\left(p^*, T^*\right) = g_{liq}\left(p^*, T^*\right).$$

It directly translates to the following relationship:

$$p^{vap}_{partial} = x^*_{vap} p^* = p_{sat}\left(T^*\right).$$

Combined with the mixture specific volume and internal energy definitions, the algebraic system to solve to determine the equilibrium state $\left(p^*, T^*, Y^*_{liq}, Y^*_{vap}\right)$, at each time step reads,

$$\begin{cases} p^{vap}_{partial} = x^*_{vap} p^* = p_{sat}\left(T^*\right) \\ v = Y^*_{liq} v_{liq}\left(p^*, T^*\right) + Y^*_{vap} v_{vap}\left(p^*, T^*\right) + \sum_{\substack{k \neq liq \\ k \neq vap}} Y_k v_k\left(p^*, T^*\right) \\ e = Y^*_{liq} e_{liq}\left(p^*, T^*\right) + Y^*_{vap} e_{vap}\left(p^*, T^*\right) + \sum_{\substack{k \neq liq \\ k \neq vap}} Y_k e_k\left(p^*, T^*\right) \\ Y^*_{vap} = 1 - Y^*_{liq} - \sum_{\substack{k \neq liq \\ k \neq vap}} Y_k \end{cases} \quad (B.1)$$

with $x^*_{vap} = \dfrac{Y_{vap}/W_{vap}}{\sum_{k \geq 4} Y_k/W_k}$.

This non-linear system can be solved using an iterative algorithm as the one given in Le Metayer et al. (2013). However, the thermochemical relaxation solver of Chiapolino et al. (2017) is preferred as it manages transitions from mixtures to pure fluids easily and is simple to implement.

The procedure is summarized as follows:

1. **Check for total evaporation ( $Y^*_{liq} \to \varepsilon$ )**

   The limit case $Y^*_{liq} \to \varepsilon$ is considered (typically $\varepsilon = 10^{-7}$). Pressure and temperature are determined using the mixture equation of state given in Section 3.2. In addition, the vapor partial pressure is calculated.

   If $T > T_{sat}\left(p^{vap}_{partial}\right)$, the solution is $Y^*_{liq} = Y_{min} = 10^{-7}$ and $Y^*_{vap} = 1 - Y_{min} - \sum_{\substack{k \neq liq \\ k \neq vap}} Y_k$.

2. **If the liquid phase "liq" is present in sufficient proportions thermodynamic equilibrium must be computed**

   Three approximate expressions of $Y^*_{vap}$ are determined:
   - One from the mixture specific volume definition:

$$Y_{Liq}^{m}(p,T) = \frac{v - \left(1 - \sum_{\substack{k \neq liq \\ k \neq vap}} Y_k\right) v_{vap}(p,T) - \sum_{\substack{k \neq liq \\ k \neq vap}} Y_k v_k(p,T)}{v_{liq}(p,T) - v_{vap}(p,T)},$$

yielding,

$$Y_{vap}^{m}(p,T) = 1 - Y_{liq}^{m}(p,T) - \sum_{\substack{k \neq liq \\ k \neq vap}} Y_k.$$

- Another one from the mixture internal energy definition:

$$Y_{liq}^{e}(p,T) = \frac{e - \left(1 - \sum_{\substack{k \neq liq \\ k \neq vap}} Y_k\right) e_{vap}(p,T) - \sum_{\substack{k \neq liq \\ k \neq vap}} Y_k e_k(p,T)}{e_{liq}(p,T) - e_{vap}(p,T)},$$

Yielding,

$$Y_{vap}^{e}(p,T) = 1 - Y_{liq}^{e}(p,T) - \sum_{\substack{k \neq liq \\ k \neq vap}} Y_k.$$

- A last one from the relationship $p_{partial}^{vap} = p_{sat}(T^*)$,

$$Y_{vap}^{sat}(p,T) = \frac{p_{sat}(T) W_{vap}}{p - p_{sat}(T)} \sum_{\substack{k > 3 \\ k \neq vap}} Y_k / W_k.$$

The approximate mass fraction $Y_{vap}^{*}$ at equilibrium that is retained is the one that produces the less variations, provided that the three associated mass transfers are of the same sign.
Let us introduce:

$$\begin{cases} r_1 = \left(Y_{vap}^{m} - Y_{vap}\right)\left(Y_{vap}^{e} - Y_{vap}\right) \\ r_2 = \left(Y_{vap}^{m} - Y_{vap}\right)\left(Y_{vap}^{sat} - Y_{vap}\right) \end{cases},$$

where $Y_{vap}$ is the initial mass fraction (from the hyperbolic step).

- If $r_1 < 0$ or $r_2 < 0$, no mass transfer is considered: $Y_{vap}^{*} = Y_{vap}$
- Otherwise the minimum mass transfer is used.

It means,

$$Y_{vap}^{*} = Y_{vap} + \text{sgn}\left[Y_{vap}^{m} - Y_{vap}\right] \times \text{Min}\left[\left|Y_{vap}^{m} - Y_{vap}\right|, \left|Y_{vap}^{e} - Y_{vap}\right|, \left|Y_{vap}^{sat} - Y_{vap}\right|\right].$$

It is clearly reminiscent of TVD type schemes and associated gradient limiters, that are converted in the present frame to source terms limiters.

As several liquids are present in SWR situations special care is needed. Phase transition of water and sodium are treated separately (in two successive steps):
- The relaxation solver is used first with liquid water and its vapor. This step is done assuming that the other gas species (sodium vapor, hydrogen and air) are frozen and thus considered as non-condensable ones. In the same way, liquid sodium and liquid soda are considered frozen.
- The relaxation solver is used secondly with liquid sodium and its vapor. The other gas species (water vapor, hydrogen and air) are frozen during this step, as well as liquid water and liquid soda.

Consequently, considering several liquids and associated vapors doesn't introduce extra difficulty.

## Appendix C. Inflow/outflow boundary condition

Tank boundary condition, with imposed pressure, temperature and mass fractions, is considered in this appendix. Under certain circumstances the inflow may become an outflow and reversely. Mathematical analysis of the mechanical and thermal equilibrium model considered in the paper shows the propagation of 3 waves at speeds: $u$, $u-c$ and $u+c$. For the sake of simplicity, we assume that the tank is located to the left of the first internal domain computational cell. Figure C.1 shows wave emission at the boundary between the tank and the first computational cell.

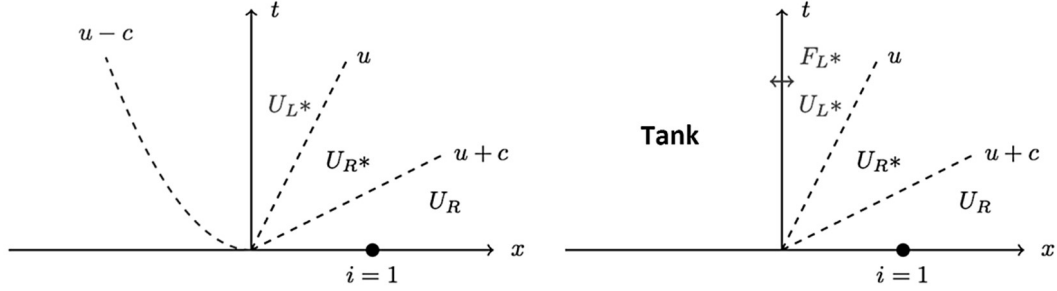

Figure C.1: Schematic representation of wave emission in the $(x,t)$ diagram for a tank inlet boundary condition

The tank being infinitely large, wave propagating in the tank at the speed $u-c$ becomes curved as a result of 3D effects. However, in multi-dimension, jump relations are unknown for such wave. Assuming quasi-steady flow between the tank state, denoted by '0' and the inlet, stagnation enthalpy and entropy invariance provide appropriate relations. However, for the present flow model the equations of state being sophisticated, isentropic transformations are approximated from the sound speed definition,

$$c_L^{2*} = c_0^2 \approx \frac{p^* - p_0}{\rho_L^* - \rho_0},$$

where c denotes the mixture sound speed.
It provides mixture density at the inlet,

$$\rho_L^* = \rho_0 + \frac{p^* - p_0}{c_0^2}.$$

In addition, the total enthalpy is conserved between the tank and the inlet section. Thus:

$$H_0 = H_L^* = e_0 + \frac{p_0}{\rho_0} = e_L^* + \frac{p^*}{\rho_L^*} + \frac{1}{2}u^{*2}.$$

Combining these two relations a function of the pressure $p^*$ is obtained:

$$e_L^*(p^*) - e_0 + \frac{p^* c_0^2}{\rho_0 c_0^2 + (p^* - p_0)} - \frac{p_0}{\rho_0} + \frac{1}{2}u^{*2}(p^*) = 0 \tag{C.1}$$

The mixture internal energy is defined as follows:

$$e = \sum_{k=1}^{N} Y_k e_k(\rho,p)$$

Using the NASG equation of state governing each phase, it is possible to re-express this definition as follows:

$$e = \overline{q} + (v(p) - \overline{b})B(p), \tag{C.2}$$

with $\overline{q} = \sum_{k=1}^{N} Y_k q_k$, $\overline{b} = \sum_{k=1}^{N} Y_k b_k$ and $B(p) = \dfrac{\sum_{k=1}^{N} Y_k c_{v,k} \dfrac{p + \gamma_k p_{\infty,k}}{p + p_{\infty,k}}}{\sum_{k=1}^{N} Y_k c_{v,k} \dfrac{\gamma_k - 1}{p + p_{\infty,k}}}$.

In addition, equation (C.1) requires the knowledge of $u^{*2}(p^*)$ which is determined by considering Rankine-Hugoniot jump relations through the wave $u+c$ propagating to the right:

$$v_R^*(p^*) = v_R + \frac{2}{p^* + p_R}\left(e_R - e_R^*(p^*)\right).$$

$$u^*(p^*) = -\sqrt{\frac{p_R - p^*}{v_R^*(p^*) - v_R}} \left( v_R^*(p^*) - v_R \right) + u_R .$$

Inserting the equation of state (C.2) into (C.1) a function depending on $p^*$ only is obtained:

$$f(p^*) = \overline{q}_L^* + \left( v_L^*(p^*) - \overline{b}_L^* \right) B(p^*) - e_0 + \frac{p^* c_0^2}{\rho_0 c_0^2 + (p^* - p_0)} - \frac{p_0}{\rho_0} + \frac{1}{2} u^{*2}(p^*) = 0 .$$

An iterative method is used for its resolution.

Under certain circumstances the inflow may become an outflow. The 'inflow' velocity $u^*(p^*)$ becomes negative. In this case the wave pattern to consider is the one of Figure C.2:

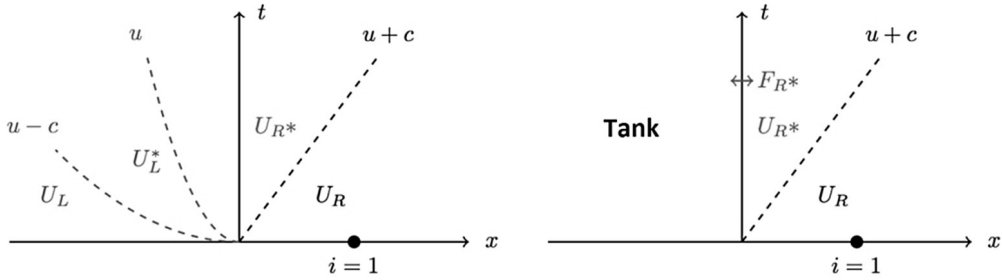

Figure C.2: Wave pattern when the inflow becomes an outflow. In that case the pressure $p^*$ is prescribed to the tank pressure $p_0$.

The specific volume in the state R* is derived from the Hugoniot relation:

$$v_R^*(p^*) = v_R + \frac{2}{p^* + p_R} \left( e_R - e_R^*(p^*) \right) ,$$

where $p^* = p_0$. The star velocity is computed with the same relation as before.